\documentclass[showpacs,preprintnumbers,amsmath,amssymb]{revtex4}

\usepackage{graphicx}
\usepackage{dcolumn}
\usepackage{bm}

\usepackage{xcolor}

\newcommand{\bq}{\begin{equation}}
\newcommand{\eq}{\end{equation}}
\newcommand{\bqa}{\begin{eqnarray}}
\newcommand{\eqa}{\end{eqnarray}}
\newcommand{\nn}{\nonumber \\}

\def\be     {\begin{equation}}
\def\ee     {\end{equation}}
\def\bea        {\begin{eqnarray}}
\def\eea        {\end{eqnarray}}
\def\bnn    {\begin{eqnarray*}}
\def\enn    {\end{eqnarray*}}

\begin{document}

\title{Nonequilibrium thermodynamics perspectives for the monotonicity of the renormalization group flow}
\author{Ki-Seok Kim$^{a,b}$ and Shinsei Ryu$^{c}$}
\affiliation{$^{a}$Department of Physics, POSTECH, Pohang, Gyeongbuk 37673, Korea \\ $^{b}$Asia Pacific Center for Theoretical Physics (APCTP), Pohang, Gyeongbuk 37673, Korea \\ $^{c}$Department of Physics, Princeton University, Princeton, New Jersey, 08540, USA}
\email[Ki-Seok Kim: ]{tkfkd@postech.ac.kr}
\email[Shinsei Ryu: ]{shinseir@princeton.edu}
\date{\today}

\begin{abstract}
  We investigate the monotonicity of the renormalization group (RG) flow
  from the perspectives of nonequilibrium thermodynamics.
  Applying the Martin-Siggia-Rose formalism to the Wilsonian RG transformation,
  %we manifest RG flow equations at the level of an effective action,
  we incorporate the RG flow equations manifestly in an effective action, 
  %(Maybe "a Wilsonian effective action" or "a Wilson-Polchinski effective action"?)
  where all coupling functions are dynamically promoted.
  As a result, we obtain an emergent holographic dual effective field theory,
  where an extra dimension appears from the Wilsonian RG transformation.
  We observe that Becchi-Rouet-Stora-Tyutin (BRST)-type transformations play an important role in the bulk effective action,
  which give rise to novel Ward identities for correlation functions between the
  renormalized coupling fields.
  As generalized fluctuation-dissipation theorems in the semiclassical
  nonequilibrium dynamics can be understood from the Ward identities of such
  BRST symmetries,
  we find essentially the same principle for the RG flow in the holographic dual
  effective field theory.
  Furthermore, we discuss how these ``nonequilibrium work identities" can be related to the monotonicity of the RG flow,
  for example, the $c-theorem$.
  In particular, we introduce an entropy functional for the dynamical coupling field 
  and show that the production rate of the total entropy functional is always positive, 
  indicating the irreversibility of the RG flow.
\end{abstract}

%\pacs{71.10.Hf, 71.30.+h, 71.10.-w, 71.10.Fd}

\maketitle

\section{Introduction}

The monotonicity of the renormalization group (RG) flow serves as 
one of the fundamental constraints 
for the dynamics of elementary degrees of freedom 
in quantum field theories 
\cite{c_theorem,a_theorem,a_f_theorem_i,f_theorem_SUSY,f_theorem_noSUSY,a_f_theorem_ii}. 
This RG monotonicity is formulated as $c-theorem$, 
where $c_{UV}$ at a UV fixed point 
should be larger than $c_{IR}$ at an IR one. 
Here, $c$ is the central charge, representing the number of degrees of freedom 
of a conformal field theory describing the corresponding fixed point. 
The $c-theorem$ states that the entanglement entropy has to decrease 
along the RG flow whatever perturbations are applied to the original fixed point 
\cite{Entanglement_Entropy_Review_RMP,a_f_theorem_EE_i,a_f_theorem_EE_ii,a_f_theorem_EE_iii,a_f_theorem_EE_iv,a_f_theorem_EE_v,c_theorem_holography_i,c_theorem_holography_ii}.

%
%{\color{red}(Optional: May be good to be more specific about dimensions explicitly.
%c in 2d, F theorem in 3d, a theorm in 4d?
%)}
%

In this study, we revisit the monotonicity of the RG flow 
from the perspectives 
of nonequilibrium thermodynamics 
\cite{MSR_Formulation_SUSY_i, MSR_Formulation_SUSY_ii, MSR_Formulation_SUSY_iii, MSR_Formulation_SUSY_iv, MSR_Formulation_SUSY_v, MSR_Formulation_SUSY_vi}. 
In this perspective, UV and IR fixed 
points can be regarded as equilibrium states while the RG flow may be 
regarded as a nonequilibrium path connecting these states.
%
%{\color{red}
%***given by a work protocol outside***.
%(I'm not sure if i understand this sentence.
%What is "outside" in the RG?)
%}
%
%{\color{red}
Viewing the RG flow as a dynamical process 
is a useful perspective, 
and explored, e.g., in Ref.\ \cite{MERA}.
%}
%**The UV and IR fixed points correspond to equilibrium states
%while the RG flow may be regarded as a nonequilibrium path 
%between two equilibrium states, given by a work protocol outside. **
%{\color{red}
In non-equilibrium statistical mechanics,
%}
it has been shown that the arrow of time in nonequilibrium dynamics can be formulated 
as generalized fluctuation-dissipation theorems 
such as Jarzynski's equality \cite{Jarzynski_i,Jarzynski_ii} and more microscopically, 
the Crooks relation \cite{Crooks_i,Crooks_ii,Crooks_iii}. 
More directly, the so-called entropy production has been shown to be responsible 
for such nonequilibrium work identities 
\cite{Entropy_Production}. 
Here, we find essentially the same principle for the RG flow and discuss how these 
``nonequilibrium work identities" 
can be related to the monotonicity of the RG flow, for example, the $c-theorem$.

We point out that generalized fluctuation-dissipation theorems or nonequilibrium work identities 
can be derived from the symmetry principle in the Schwinger-Keldysh path integral formulation 
\cite{Schwinger_Keldysh_Symmetries_i, Schwinger_Keldysh_Symmetries_ii, Schwinger_Keldysh_Symmetries_iii, Schwinger_Keldysh_Symmetries_iv, Schwinger_Keldysh_Symmetries_v, Schwinger_Keldysh_Symmetries_vi}. 
Here, the number of elementary degrees of freedom 
is doubled to cause some redundancies in the path integral description. 
As a result, certain topological symmetries involved with unitarity appear to be described 
by two types of Becchi-Rouet-Stora-Tyutin (BRST) symmetries. 
In addition to these topological symmetries, 
there are microscopic time-reversal symmetries referred to as Kubo-Martin-Schwinger (KMS) ones 
if the initial state is in thermal equilibrium. 
These KMS symmetries can be described by two additional fermion-type symmetries. 
It turns out that these four types of fermion symmetries form an extended $\mathcal{N} = 2$ 
equivariant cohomology algebra \cite{TQFT_Witten_Type}. 
This $\mathcal{N} = 2$ supersymmetry gives strong constraints to the nonequilibrium thermodynamics.

To avoid any possible confusion and uncertainties, 
we limit ourselves to the semiclassical nonequilibrium dynamics, for example, Langevin-type dynamics. 
Then, the Schwinger-Keldysh formulation is reduced into the Martin-Siggia-Rose (MSR) formalism 
\cite{MSR_Formulation_i, MSR_Formulation_ii, MSR_Formulation_iii} 
for the description of a nonequilibrium protocol with two boundary conditions 
(equilibrium states) \cite{NEQ_textbook}. 
In this case, the four kinds of BRST symmetries have been more well-established to give Ward identities 
for correlation functions 
\cite{MSR_Formulation_SUSY_i, MSR_Formulation_SUSY_ii, MSR_Formulation_SUSY_iii, MSR_Formulation_SUSY_iv, MSR_Formulation_SUSY_v}. 
These Ward identities can be translated as nonequilibrium work identities, i.e., Jarzynski's and Crooks' identities.

In the present study, we apply this strategy to the Wilsonian RG transformation. 
Applying the MSR formalism to the Wilsonian RG transformation, 
%we manifest RG flow equations at the level of an effective action, 
we incorporate the RG flow equations manifestly in an effective action,
%(Maybe "a Wilsonian effective action" or "a Wilson-Polchinski effective action"?)
where all coupling functions are dynamically promoted to be coupling fields. 
As a result, we obtain an emergent holographic dual effective field theory, where 
%one additional 
an extra dimension appears from the Wilsonian RG transformation 
\cite{Nonperturbative_Wilson_RG_Disorder,Nonperturbative_Wilson_RG,Einstein_Klein_Gordon_RG_Kim,Einstein_Dirac_RG_Kim,RG_GR_Geometry_I_Kim,RG_GR_Geometry_II_Kim,Kondo_Holography_Kim,
Kitaev_Entanglement_Entropy_Kim,RG_Holography_First_Kim,Emergent_AdS2_BH_RG}. 
Here, the holography is realized by the appearance of the extra dimension, identified with an RG scale, 
and the duality indicates that the effective field theory 
is written in terms of collective order parameter fields 
instead of the original fields as the Landau-Ginzburg free energy functional.
We furthermore turn on irrelevant perturbations at the UV scale, 
which plays the role of noise in the language of stochastic dynamics. 
We observe that there exist four kinds of BRST-type emergent symmetries in the bulk effective action, 
which give rise to novel Ward identities for correlation functions between the renormalized coupling fields. 
As a result, we find a generalized fluctuation-dissipation theorem for the RG flow, 
where the standard form of the theorem is modified by the RG transformation. 
Based on this thermodynamics perspective, we discuss the monotonicity of the RG flow, 
introducing an effective entropy functional in terms of the coupling field. 
It turns out that the rate of the total entropy functional is always positive, 
indicating the irreversibility of the RG flow. 
This indicates how the generalized fluctuation-dissipation theorem can be related to the monotonicity of the RG flow, 
for example, the $c-theorem$.

\section{A review on stochastic thermodynamics in the Langevin system} 

Since the main objective of the present study is to reformulate the monotonicity or irreversibility of the RG flow 
from the stochastic thermodynamics perspective, 
it would be helpful to review some mathematical constructions for the stochastic thermodynamics 
\cite{MSR_Formulation_SUSY_i, MSR_Formulation_SUSY_ii, MSR_Formulation_SUSY_iii, MSR_Formulation_SUSY_iv, MSR_Formulation_SUSY_v, MSR_Formulation_SUSY_vi, Entropy_Production, MSR_Formulation_i, MSR_Formulation_ii,MSR_Formulation_iii}. 
As a prototypical example, we consider the overdamped dynamics of a particle in one dimension subject to a force, 
described by the Langevin equation,
\bqa && \partial_{t} x(t) = \mu F(x(t),\lambda(t)) + \xi(t) . \eqa
Here, $ F(x(t),\lambda(t)) = - \partial_{x} V(x(t),\lambda(t)) + f(x(t),\lambda(t))$ is the force, 
where $V(x(t),\lambda(t))$ is a conservative potential and $f(x(t),\lambda(t))$ is an external force. 
These force sources may be time-dependent through an external control parameter $\lambda(t)$ 
varied according to some prescribed experimental 
protocol from $\lambda(0) = \lambda_{0}$ to $\lambda(t_{f}) = \lambda_{f}$. 
$\mu$ is the mobility of the particle. 
$\xi(t)$ serves as stochastic increments modeled as Gaussian white noise,
\bqa && \langle \xi(t) \xi(t') \rangle = 2 D \delta(t-t') , \label{White_Noise} \eqa
where $D$ is the diffusion constant, given by the Einstein relation $D = \beta^{-1} \mu$ 
at temperature $T = \beta^{-1}$ in equilibrium.

To investigate the symmetries of the Langevin equation, it is more convenient to consider a generating functional 
for physical observables, 
analogous to the partition function. Here, the following identity is essential,
\bqa && 1 = \int_{x_{i}}^{x_{f}} D x(t) \delta\Big(\partial_{t} x(t) - \mu F(x(t),\lambda(t)) - \xi(t)\Big) \mbox{det}\Big(\partial_{t} - \mu \partial_{x} F(x(t),\lambda(t))\Big) . \eqa
This $\delta$-function identity is nothing but the Faddeev-Popov procedure 
for the path integral quantization of gauge fields \cite{QFT_textbook}. 
As a result, one may propose a generating functional for the overdamped Langevin dynamics subject to a force as follows 
\cite{MSR_Formulation_SUSY_i, MSR_Formulation_SUSY_ii,MSR_Formulation_SUSY_iii, MSR_Formulation_SUSY_iv, MSR_Formulation_SUSY_v, MSR_Formulation_SUSY_vi, MSR_Formulation_i, MSR_Formulation_ii,MSR_Formulation_iii}
\bqa && \mathcal{W} = \mathcal{N} \int_{x_{i}}^{x_{f}} D x(t) D p(t) D c(t) D \bar{c}(t) \int D \xi(t) \exp\Big( - \frac{1}{4 D} \int_{t_{i}}^{t_{f}} d t \xi^{2}(t) \Big) 
\nn && \times \exp\Big[- \int_{t_{i}}^{t_{f}} d t \Big\{ i p(t) \Big( \partial_{t} x(t) - \mu F(x(t),\lambda(t)) - \xi(t) \Big) + \bar{c}(t) \Big(\partial_{t} - \mu \partial_{x} F(x(t),\lambda(t))\Big) c(t) \Big\} \Big] . \eqa
Here, $p(t)$ is a Lagrange multiplier, identified with a canonical momentum to the position $x(t)$, 
and $c(t)$ is a fermion variable to take the Jacobian factor with its canonical conjugate partner $\bar{c}(t)$.
$\mathcal{N}$ is a normalization constant to reproduce Eq.\ (\ref{White_Noise}). 
Performing the average with respect to random noise fluctuations, we obtain 
\begin{small}
\bqa && \mathcal{W} = \mathcal{N} \int_{x_{i}}^{x_{f}} D x(t) D p(t) D c(t) D \bar{c}(t) \exp\Big[- \int_{t_{i}}^{t_{f}} d t \Big\{ i p(t) \Big( \partial_{t} x(t) - \mu F(x(t),\lambda(t)) \Big) + D p^{2}(t) + \bar{c}(t) \Big(\partial_{t} - \mu \partial_{x} F(x(t),\lambda(t))\Big) c(t) \Big\} \Big] . \nn \label{Generating_Functional_Langevin} \eqa
\end{small}
Based on this path integral formulation, Refs.\ \cite{MSR_Formulation_SUSY_i,MSR_Formulation_SUSY_ii,MSR_Formulation_SUSY_iii,MSR_Formulation_SUSY_iv,MSR_Formulation_SUSY_v,MSR_Formulation_SUSY_vi} investigated BRST symmetries and discussed Ward identities. 
In this study, we apply this framework to the RG flow and reveal symmetries of the RG flow.

One important ingredient involved with the monotonicity of the RG flow is entropy production in the Langevin system. 
Introducing the following probability distribution,
\bqa && p(x,t) = \langle \delta(x-x(t)) \rangle = \mathcal{N} \int D \xi(t') \exp\Big( - \frac{1}{4 D} \int_{t_{i}}^{t} d t' \xi^{2}(t') \Big) \delta(x-x(t)) , \eqa 
where the average of random noise fluctuations is taken. 
Then, one obtains the Fokker-Planck equation for the probability distribution function to find the particle at $x$ 
and at time $t$, 
\bqa && \partial_{t} p(x,t) = - \partial_{x} j(x,t) = - \partial_{x} [ (\mu F(x,\lambda) - D \partial_{x}) p(x,t) ] . \eqa
$j(x,t) = (\mu F(x,\lambda) - D \partial_{x}) p(x,t)$ is the conserved current. 
This partial differential equation must be augmented by a normalized initial distribution, $p(x,0) = p_{0}(x)$. 
In Appendix A, we show our intuitive derivation for this Fokker-Planck equation. 
It is straightforward to see the formal path integral expression for the probability distribution function as follows
\bqa && p(x,t) = \frac{\mathcal{N}}{\mathcal{W}} \int D \xi(t') \exp\Big( - \frac{1}{4 D} \int_{t_{i}}^{t} d t' \xi^{2}(t') \Big) \int_{x_{i}}^{x} D x(t') D p(t') D \bar{c}(t') D c(t') 
\nn && \quad \times \exp\Big[- \int_{t_{i}}^{t} d t' \Big\{ i p(t') \Big( \partial_{t'} x(t') - \mu F(x(t'),\lambda(t')) - \xi(t') \Big) + \bar{c}(t') \Big(\partial_{t'} - \mu \partial_{x} F(x(t'),\lambda(t'))\Big) c(t') \Big\} \Big] , \eqa
where the normalization constant or the generating functional is given by Eq.\ (\ref{Generating_Functional_Langevin}).
%
%\bqa && \mathcal{W}(t) = \int_{x_{i}}^{x_{f}} D x(t') D p(t') D \bar{c}(t') D c(t') \int D \xi(t') \exp\Big( - \frac{1}{4 D} \int_{t_{i}}^{t_{f}} d t' \xi^{2}(t') \Big) \nn && \exp\Big[- \int_{t_{i}}^{t_{f}} d t' \Big\{ %i p(t') \Big( \partial_{t'} x(t') - \mu F(x(t'),\lambda(t')) - \xi(t') \Big) + \bar{c}(t') \Big(\partial_{t'} - \mu \partial_{x} F(x(t'),\lambda(t'))\Big) c(t') \Big\} \Big] . \eqa
%
One can verify
\bqa && \int_{x_{i}}^{x_{f}} d x p(x,t) = 1 . \eqa

To discuss the entropy production in the forced overdamped Langevin dynamics, 
Ref.\ \cite{Entropy_Production} proposed a trajectory-dependent entropy for the particle or system as
\bqa && s_{sys}(x,t) = - \ln p(x,t) . \eqa
This definition is consistent with the common definition of a nonequilibrium Gibbs entropy, given by
\bqa && S_{sys}(t) = \langle s_{sys}(x,t) \rangle= - \int_{x_{i}}^{x_{f}} d x p(x,t) \ln p(x,t) . \eqa
This microscopic entropy gives rise to the macroscopic thermodynamic entropy 
for an equilibrium Boltzmann distribution at fixed $\lambda$, 
\bqa && s_{sys}(x,t) = \beta [V(x,\lambda) - F(\lambda)] , \eqa
where the equilibrium free energy $F(\lambda)$ is $F(\lambda) = - \beta^{-1} \ln \int_{x_{i}}^{x_{f}} d x e^{- \beta V(x,\lambda)}$ with the conserved potential $V(x,\lambda)$ introduced before. 
Then, it is natural to consider the rate of heat dissipation in the environment as 
\bqa && \partial_{t} q(x,t) = F(x,\lambda) \partial_{t} x(t) = \beta^{-1} \partial_{t} s_{env}(x,t) . \eqa
Accordingly, one may identify the exchanged heat with an increase 
in the environment entropy $s_{env}(x,t)$ at temperature $\beta^{-1} = D / \mu$. 

Combining these two contributions, Ref.\ \cite{Entropy_Production} 
found the trajectory-dependent total entropy production rate as follows
\bqa && \partial_{t} s_{tot}(x,t) = \partial_{t} s_{env}(x,t) + \partial_{t} s_{sys}(x,t) = \frac{\partial_{x} j(x,t)}{p(x,t)} + \frac{j(x,t)}{D p(x,t)} \partial_{t} x(t) . \eqa
Taking the ensemble average, Ref.\ \cite{Entropy_Production} showed that 
the averaged total entropy production rate is always positive, given by
\bqa && \partial_{t} S_{tot}(t) = \langle \partial_{t} s_{tot}(x,t) \rangle = \int_{x_{i}}^{x_{f}} d x \frac{j^{2}(x,t)}{D p(x,t)} \geq 0 , \eqa 
where the equality holds in equilibrium only. The ensemble-averaged entropy production rate of the environment is given by
\bqa && \partial_{t} S_{env}(x,t) = \langle \partial_{t} s_{env}(x,t) \rangle = \beta \int_{x_{i}}^{x_{f}} d x F(x,t) j(x,t) , \eqa
where the force $F(x,t)$ and the conserved current $j(x,t)$ have been introduced above. 
In this study, we discuss the entropy production of the RG flow, following this line of thought.

\section{To manifest the renormalization group flow in the level of an effective action} 

\subsection{Wilsonian renormalization group transformation} 

We consider a partition function as follows
\bqa && Z(\Lambda_{uv}) = \int D \psi_{\sigma}(x;\Lambda_{uv}) \exp\Big\{ - \int d^{D} x \mathcal{L}[\psi_{\sigma}(x;\Lambda_{uv});\{\lambda_{a}(\Lambda_{uv})\};\Lambda_{uv}] \Big\} . \eqa 
Here, $\Lambda_{uv}$ is a UV cutoff, 
where the corresponding effective Lagrangian 
$\mathcal{L}[\psi_{\sigma}(x;\Lambda_{uv});\{\lambda_{a}(\Lambda_{uv})\};\Lambda_{uv}]$ is defined. 
$\psi_{\sigma}(x;\Lambda_{uv})$ is a dynamical matter field at a given spacetime $x$, 
where $\sigma$ denotes its flavor index $\sigma = 1, \ldots, N$. 
$\{\lambda_{a}(\Lambda_{uv})\}$ represents a set of coupling functions 
such as velocity, interaction coefficients, etc., denoted by the subscript $a$. 

Performing the Wilsonian RG transformation, 
we obtain the following expression for the partition function
\bqa && Z(z_{f}) = \int D \psi_{\sigma}(x;z_{f}) \exp\Big\{ - \int d^{D} x \Big( \mathcal{L}[\psi_{\sigma}(x;z_{f});\{\lambda_{a}(x,z_{f})\};z_{f}] + N \int_{\Lambda_{uv}}^{z_{f}} d z \mathcal{V}_{rg}[\{\lambda_{a}(x,z)\};z] \Big) \Big\} , \eqa
where the UV cutoff $\Lambda_{uv}$ is lowered to be $z_{f}$. 
In other words, all the dynamical fields $\psi_{\sigma}(x;z_{f})$ 
and all the coupling functions $\lambda_{a}(x,z_{f})$ are defined at a lower cutoff $z_{f}$, 
where the dynamical fields between $z_{f}$ and $\Lambda_{uv}$ 
are integrated over to introduce an effective potential 
$N \int_{\Lambda_{uv}}^{z_{f}} d z \mathcal{V}_{rg}[\{\lambda_{a}(x,z)\};z]$ 
into the partition function 
\cite{Nonperturbative_Wilson_RG_Disorder,Nonperturbative_Wilson_RG,Einstein_Klein_Gordon_RG_Kim}. 
%{\bf (SR: perhaps we need to cite some papers here.)}

Considering that the partition function is invariant under the RG transformation, 
regardless of the cutoff scale, we observe that the effective potential is 
\bqa && \mathcal{V}_{rg}[\{\lambda_{a}(x,z)\};z] = - \frac{1}{N} \ln \int_{\Lambda(z)} D \psi_{\sigma}(x;z) \exp\Big\{ - \int d^{D} x \mathcal{L}[\psi_{\sigma}(x;z);\{\lambda_{a}(x,z)\};z] \Big\} , \eqa
at a given scale $z$. 
Accordingly, all the coupling functions are renormalized to be
\bqa && \frac{\partial \lambda_{a}(x,z)}{\partial z} = \beta_{a}[\{\lambda_{a}(x,z)\};z] . \eqa
Here, the RG $\beta$-function for a coupling function $\lambda_{a}(x,z)$ 
is given by the first-order derivative of the effective potential 
with respect to $\lambda_{a}(x,z)$ as follows
\bqa && \beta_{a}[\{\lambda_{a}(x,z)\};z] = - \frac{\partial \mathcal{V}_{rg}[\{\lambda_{a}(x,z)\};z]}{\partial \lambda_{a}(x,z)} . \label{RG_beta_ft_general} \eqa
Implementing this calculation explicitly, 
we see that the RG $\beta$-function is given by a renormalized vertex function,
\bqa && N \beta_{a}[\{\lambda_{a}(x,z)\};z] \nn && = \frac{1}{Z(z)} \int_{\Lambda(z)} D \psi_{\sigma}(x;z) \Bigg( \frac{\partial }{\partial \lambda_{a}(x,z)} \int d^{D} x \mathcal{L}[\psi_{\sigma}(x;z);\{\lambda_{a}(x,z)\};z] \Bigg) \exp\Big\{ - \int d^{D} x \mathcal{L}[\psi_{\sigma}(x;z);\{\lambda_{a}(x,z)\};z] \Big\} , \nn \eqa
where $Z(z)$ is an effective partition function at a given scale $z$,
\bqa && Z(z) = \int_{\Lambda(z)} D \psi_{\sigma}(x;z) \exp\Big\{ - \int d^{D} x \mathcal{L}[\psi_{\sigma}(x;z);\{\lambda_{a}(x,z)\};z] \Big\} . \eqa

\subsection{To manifest the renormalization group flow in the level of an effective action}

To manifest the RG flow at the level of an effective action, 
we consider the following identity
\bqa && 1 = \int D \lambda_{a}(x,z) \delta \Big( \partial_{z} \lambda_{a}(x,z) - \beta_{a}[\{\lambda_{a}(x,z)\};z] \Big) \mbox{det} \Big(\partial_{z} \delta_{ab} - \frac{\partial \beta_{a}[\{\lambda_{a}(x,z)\};z]}{\partial \lambda_{b}(x,z)} \Big) . \eqa
Here, $\mbox{det} \Big(\partial_{z} \delta_{ab} - \frac{\partial \beta_{a}[\{\lambda_{a}(x,z)\};z]}{\partial \lambda_{b}(x,z)} \Big)$ 
may be regarded as a Jacobian factor for the functional integral. 
Introducing this $\delta$-function identity into the partition function, we obtain
\bqa && Z(z_{f}) = \int D \psi_{\sigma}(x,z_{f}) D \lambda_{a}(x,z) \delta \Big( \partial_{z} \lambda_{a}(x,z) - \beta_{a}[\{\lambda_{a}(x,z)\};z] \Big) \mbox{det} \Big(\partial_{z} \delta_{ab} - \frac{\partial \beta_{a}[\{\lambda_{a}(x,z)\};z]}{\partial \lambda_{b}(x,z)} \Big) \nn && \exp\Big\{ - \int d^{D} x \Big( \mathcal{L}[\psi_{\sigma}(x,z_{f});\{\lambda_{a}(x,z_{f})\};z_{f}] + N \int_{\Lambda_{uv}}^{z_{f}} d z \mathcal{V}_{rg}[\{\lambda_{a}(x,z)\};z] \Big) \Big\} . \eqa
Now, the coupling function is promoted to be a dynamical coupling field, 
which appears as the path integral formulation with the $\delta$-function constraint. 
This is essentially the same as the Faddeev-Popov procedure 
for the path integral quantization of gauge fields \cite{QFT_textbook}, 
also applied to the semiclassical nonequilibrium physics, for example, 
the path integral formulation of Langevin dynamics, 
and referred to as the MSR formalism \cite{MSR_Formulation_i, MSR_Formulation_ii, MSR_Formulation_iii} discussed before. 
Here, the RG flow corresponds to the Langevin equation.

It is straightforward to exponentiate the $\delta$-function constraint as follows
\bqa && Z(z_{f}) = \int D \psi_{\sigma}(x,z_{f}) D \lambda_{a}(x,z) D \pi_{a}(x,z) D \bar{c}_{a}(x,z) D c_{a}(x,z) \exp\Big[ - \int d^{D} x \mathcal{L}[\psi_{\sigma}(x,z_{f});\{\lambda_{a}(x,z_{f})\};z_{f}] \nn && - N \int_{\Lambda_{uv}}^{z_{f}} d z \int d^{D} x \Big\{ \pi_{a}(x,z) \Big( \partial_{z} \lambda_{a}(x,z) - \beta_{a}[\{\lambda_{a}(x,z)\};z] \Big) + \bar{c}_{a}(x,z) \Big(\partial_{z} \delta_{ab} - \frac{\partial \beta_{a}[\{\lambda_{a}(x,z)\};z]}{\partial \lambda_{b}(x,z)} \Big) c_{b}(x,z) \nn && + \mathcal{V}_{rg}[\{\lambda_{a}(x,z)\};z] \Big\} \Big] . \label{HDEFT_No_Dynamics} \eqa  
$\pi_{a}(x,z)$ is a Lagrange multiplier field to impose the RG flow constraint, 
which corresponds to the canonical momentum of the coupling field $\lambda_{a}(x,z)$. 
In the Schwinger-Keldysh formulation, 
it is identified with a quantum field denoted by the subscript $a$ or $qu$ 
in the standard notation. 
$c_{a}(x,z)$ ($\bar{c}_{a}(x,z)$) is an auxiliary fermion field to 
take care of the Jacobian factor, 
referred to as the Faddeev-Popov ghost. 
$z$ is an RG scale, which serves as a cutoff scale for the Wilsonian RG transformation. 
Interestingly, this RG scale plays the role of an extra dimension, 
which reminds us of the holographic duality conjecture \cite{Holographic_Duality_I,Holographic_Duality_II,Holographic_Duality_III,Holographic_Duality_IV,Holographic_Duality_V,Holographic_Duality_VI,Holographic_Duality_VII}, 
where $\mathcal{S}_{eff}[\{\pi_{a}(x,z),\lambda_{a}(x,)\},\{\bar{c}_{a}(x,z),c_{a}(x,z)\};z_{f},\Lambda_{uv}] = N \int_{\Lambda_{uv}}^{z_{f}} d z \int d^{D} x \Big\{ \pi_{a}(x,z) \Big( \partial_{z} \lambda_{a}(x,z) - \beta_{a}[\{\lambda_{a}(x,z)\};z] \Big) + \bar{c}_{a}(x,z) \Big(\partial_{z} \delta_{ab} - \frac{\partial \beta_{a}[\{\lambda_{a}(x,z)\};z]}{\partial \lambda_{b}(x,z)} \Big) c_{b}(x,z) + \mathcal{V}_{rg}[\{\lambda_{a}(x,z)\};z] \Big\}$ 
corresponds to an effective bulk action, 
supported by an effective boundary action of $\int d^{D} x \mathcal{L}[\psi_{\sigma}(x,z_{f});\{\lambda_{a}(x,z_{f})\};z_{f}]$. 
Here, the duality means that the bulk effective action is written 
in terms of the coupling fields $\{\lambda_{a}(x,z)\}$, 
regarded to as collective dual fields to the corresponding matter composites. 
In other words, 
$\lambda_{a}(x,z)$ 
is dual to $\frac{\partial \mathcal{L}[\psi_{\sigma}(x;z);\{\lambda_{a}(x,z)\};z]}{\partial \lambda_{a}(x,z)}$.

Before going further, we point out an essential approximation in this effective field theory.
First of all, nonlocal terms are neglected in the resulting effective action that manifests the RG flows of the coupling fields. 
We recall that the Wilsonian RG transformation generates nonlocal terms inevitably. 
Here, the RG $\beta$-function Eq.\ (\ref{RG_beta_ft_general}) is given by a Green's function of the corresponding matter field 
at a given energy scale $z$. 
Since the Green's function is bi-local, i.e., depending on $x$ and $x'$, 
nonlocality is unavoidable. 
Such emergent nonlocal interactions can however be ``localized" 
at the cost of introducing higher-spin fields to decompose them in a local fashion based on the corresponding group structure 
\cite{Higher_Spin_Gauge_Theory_I, Higher_Spin_Gauge_Theory_II, Higher_Spin_Gauge_Theory_III, Higher_Spin_Gauge_Theory_IV, Holography_Higher_Spin_RG_I, Holography_Higher_Spin_RG_II, Holography_Higher_Spin_RG_III, Higher_Spin_Gauge_Theory_V}. 
In other words, integrating over such higher-spin fields can give rise to an effective gravity theory including only up to the spin two fields, 
but in the presence of effective nonlocal interactions between gravitons. 
In most cases, we will work with a proper local truncation of these RG-generated nonlocal terms, 
keeping the original form of the effective Lagrangian as in the conventional RG transformation 
\cite{Nonperturbative_Wilson_RG_Disorder, Nonperturbative_Wilson_RG, Einstein_Klein_Gordon_RG_Kim, Einstein_Dirac_RG_Kim, RG_GR_Geometry_I_Kim, RG_GR_Geometry_II_Kim, Kondo_Holography_Kim,
Kitaev_Entanglement_Entropy_Kim,RG_Holography_First_Kim,Emergent_AdS2_BH_RG}. 
Here, based on the gradient expansion in the limit of $\Delta x = x - x' \rightarrow 0$, 
we have only local terms in the resulting effective action. 
This issue was well summarized in Ref.\ \cite{Einstein_Klein_Gordon_RG_Kim}. 

\subsection{To promote coupling functions to dynamical fields: Irrelevant deformations}

Although the above reformulation for the Wilsonian RG transformation 
is rather analogous to the holographic dual effective field theory, 
there exists one important difference: 
The coupling field $\lambda_{a}(x,z)$ is not fully dynamical, 
whose dynamics is semiclassical, given by the RG flow equation. 
To promote the coupling field to be fully dynamical, 
we consider the following UV deformation,
\begin{small}
\bqa && Z(\Lambda_{uv}) = \int D \psi_{\sigma}(x;\Lambda_{uv}) D \lambda_{a}(x;\Lambda_{uv}) \exp\Big\{ - \int d^{D} x \Big( \mathcal{L}[\psi_{\sigma}(x;\Lambda_{uv});\{\lambda_{a}(x;\Lambda_{uv})\};\Lambda_{uv}] + \frac{1}{2 \Gamma_{a}}[\lambda_{a}(x;\Lambda_{uv}) - \bar{\lambda}_{a}(\Lambda_{uv})]^{2} \Big) \Big\} . \nn \eqa 
\end{small}
Here, we introduced random fluctuations of the coupling fields at UV, 
where $\Gamma_{a}$ denotes the variance around 
the mean value $\bar{\lambda}_{a}(\Lambda_{uv})$. 
Taking the $\Gamma_{a} \rightarrow 0$ limit, 
we recover the previous formulation, 
where $\lambda_{a}(\Lambda_{uv})$ is replaced with $\bar{\lambda}_{a}(\Lambda_{uv})$.
This UV deformation can be thought of as 
averaging over theories with different coupling constants \cite{Nonperturbative_Wilson_RG_Disorder}
or turning on some irrelevant perturbations akin to the $T\bar{T}$ deformation \cite{TTbar_Deformation}. 
%To have the former interpretation, strictly speaking, 
%we may have to use the replica trick? I'm not sure..  
%We can also cite some papers here.

To understand the physical meaning of this UV deformation more precisely, 
we perform the Gaussian integral with respect to $\lambda_{a}(x;\Lambda_{uv})$. 
Then, we obtain
\begin{small}
\bqa && Z(\Lambda_{uv}) = \int D \psi_{\sigma}(x;\Lambda_{uv}) \exp\Big[ - \int d^{D} x \Big\{ \mathcal{L}[\psi_{\sigma}(x;\Lambda_{uv});\{\bar{\lambda}_{a}(\Lambda_{uv})\};\Lambda_{uv}] + \frac{\Gamma_{a}}{2} \Big(\frac{\partial \mathcal{L}[\psi_{\sigma}(x;\Lambda_{uv});\{\lambda_{a}(x;\Lambda_{uv})\};\Lambda_{uv}]}{\partial \lambda_{a}(x;\Lambda_{uv})}\Big)^{2} \Big\} \Big] . \nn \eqa 
\end{small}
Suppose the Gross-Neveu model for spontaneous chiral symmetry breaking 
as $\mathcal{L}[\psi_{\sigma}(x;\Lambda_{uv});\{\lambda_{a}(x;\Lambda_{uv})\};\Lambda_{uv}] = \bar{\psi}_{\sigma}(x;\Lambda_{uv}) i \gamma^{\mu} \partial_{\mu} \psi_{\sigma}(x;\Lambda_{uv}) + \frac{\lambda_{\chi}(x;\Lambda_{uv})}{2} \bar{\psi}_{\sigma}(x;\Lambda_{uv}) \psi_{\sigma}(x;\Lambda_{uv}) \bar{\psi}_{\sigma'}(x;\Lambda_{uv}) \psi_{\sigma'}(x;\Lambda_{uv})$. 
Then, the last term is $\Big(\frac{\partial \mathcal{L}[\psi_{\sigma}(x;\Lambda_{uv});\{\lambda_{a}(x;\Lambda_{uv})\};\Lambda_{uv}]}{\partial \lambda_{a}(x;\Lambda_{uv})}\Big)^{2} \sim [\bar{\psi}_{\sigma}(x;\Lambda_{uv}) \psi_{\sigma}(x;\Lambda_{uv})]^{4}$ \cite{Emergent_AdS2_BH_RG}, 
generally irrelevant at the Gaussian fixed point and expected not to change the RG flow 
as long as weak $\Gamma_{\chi}$ is concerned.
Considering these random fluctuations of the coupling functions at UV and 
performing the Wilsonian RG transformation, 
we obtain \cite{Nonperturbative_Wilson_RG_Disorder, Nonperturbative_Wilson_RG, Einstein_Klein_Gordon_RG_Kim,Emergent_AdS2_BH_RG}
\bqa && Z(z_{f}) = \int D \psi_{\sigma}(x,z_{f}) D \lambda_{a}(x,z) D \pi_{a}(x,z) D \bar{c}_{a}(x,z) D c_{a}(x,z) \nn && \exp\Big[ - \int d^{D} x \Big( \mathcal{L}[\psi_{\sigma}(x,z_{f});\{\lambda_{a}(x,z_{f})\};z_{f}] + \frac{1}{2 \Gamma_{a}}[\lambda_{a}(x,\Lambda_{uv}) - \bar{\lambda}_{a}(\Lambda_{uv})]^{2} \Big) \nn && - N \int_{\Lambda_{uv}}^{z_{f}} d z \int d^{D} x \Big\{ \pi_{a}(x,z) \Big( \partial_{z} \lambda_{a}(x,z) - \beta_{a}[\{\lambda_{a}(x,z)\};z] \Big) - \frac{\Gamma_{a}}{2} \pi_{a}^{2}(x,z) \nn && + \bar{c}_{a}(x,z) \Big(\partial_{z} \delta_{ab} - \frac{\partial \beta_{a}[\{\lambda_{a}(x,z)\};z]}{\partial \lambda_{b}(x,z)} \Big) c_{b}(x,z) + \mathcal{V}_{rg}[\{\lambda_{a}(x,z)\};z] \Big\} \Big] , \label{HDEFT_Dynamics} \eqa  
where $- \frac{\Gamma_{a}}{2} \pi_{a}^{2}(x,z)$ 
appeared to give the dynamics to $\lambda_{a}(x,z)$ in the extradimension. 
It is trivial to check out that the $\Gamma_{a} \rightarrow 0$ limit reproduces 
Eq.\ (\ref{HDEFT_No_Dynamics}). 

In this study, we claim that the holographic dual effective field theory 
[Eq.\ (\ref{HDEFT_Dynamics})] enjoys essentially the same structure 
as the MSR formalism of the Langevin dynamics except for the following two aspects: 
One is the presence of the effective potential $\mathcal{V}_{rg}[\{\lambda_{a}(x,z)\};z]$, 
which arises from quantum fluctuations in the Wilsonian RG transformation, 
and the other is the existence of the boundary action, 
which defines both UV and IR boundary conditions for the bulk effective action. 
%
%Below, we show that the existence of the effective potential is responsible for the monotonicity or irreversibility of the RG flow, which breaks emergent BRST-type symmetries.
%
In spite of these two different aspects, 
we derive a generalized fluctuation-dissipation theorem 
for the RG flow and show the monotonicity of the RG flow to hold as the Langevin dynamics, 
where the existence of the effective potential gives rise to some corrections in the formulae.

\section{Emergent BRST ``symmetries" in the RG flow}

\subsection{Four types of BRST transformations}

Following the MSR formalism of the Langevin dynamics, 
we investigate emergent BRST symmetries of the RG-flow manifested effective field theory,
\begin{small}
\bqa && Z(z_{f}) = \int D \psi_{\sigma}(x,z_{f}) D \lambda(x,z) D \pi(x,z) D \bar{c}(x,z) D c(x,z) \exp\Big[ - \int d^{D} x \Big( \mathcal{L}[\psi_{\sigma}(x,z_{f});\lambda(x,z_{f});z_{f}] + \frac{N}{2 \Gamma}[\lambda(x,\Lambda_{uv}) - \bar{\lambda}(\Lambda_{uv})]^{2} \Big) \nn && - N \int_{\Lambda_{uv}}^{z_{f}} d z \int d^{D} x \Big\{ \pi(x,z) \Big( \partial_{z} \lambda(x,z) - \beta[\lambda(x,z);z] \Big) - \frac{\Gamma}{2} \pi^{2}(x,z) + \bar{c}(x,z) \Big(\partial_{z} - \frac{\partial \beta[\lambda(x,z);z]}{\partial \lambda(x,z)} \Big) c(x,z) + \mathcal{V}_{rg}[\lambda(x,z);z] \Big\} \Big] . \eqa  
\end{small}
Here, we considered the case of one coupling field for simplicity.

One may consider four types of BRST transformations 
in this holographic dual effective field theory 
as the case of the Langevin dynamics \cite{MSR_Formulation_SUSY_i,MSR_Formulation_SUSY_ii,MSR_Formulation_SUSY_iii,MSR_Formulation_SUSY_iv,MSR_Formulation_SUSY_v,MSR_Formulation_SUSY_vi}. 
We recall that in the Schwinger-Keldysh formulation, 
the first two BRST symmetries with their charges $Q$ and $\bar{Q}$ are topological in origin, 
related with the unitarity. 
These two BRST symmetries do not commute with the KMS ones 
\cite{Schwinger_Keldysh_Symmetries_vi}. 
Considering both the BRST and KMS symmetries, 
we have to introduce additional two fermion-type symmetries 
with their charges $D$ and $\bar{D}$. 
Although $D$ and $\bar{D}$ correspond to the superderivatives 
in the superspace formulation as discussed in Appendix B, 
we also call these additional fermionic symmetries BRST-type symmetries. 
The first two BRST transformations lead 
the bulk kinetic energy $\pi(x,z) \Big( \partial_{z} \lambda(x,z) - \beta[\lambda(x,z);z] \Big) - \frac{\Gamma}{2} \pi^{2}(x,z) + \bar{c}(x,z) \Big(\partial_{z} - \frac{\partial \beta[\lambda(x,z);z]}{\partial \lambda(x,z)} \Big) c(x,z)$ 
to be invariant while the last two do not. 
The effective potential $\mathcal{V}_{rg}[\lambda(x,z);z]$ 
does transform under all these BRST transformations. 
As a result, there do not exist any BRST-type emergent symmetries in this RG flow, 
precisely speaking. 
However, such BRST noninvariant terms are expressed in a ``universal" way. 
As a result, we can derive generalized Ward identities from these four types of 
BRST transformations and find some constraints for correlation functions of the coupling field.

The first BRST transformation is given by
\bqa && \delta_{Q} \lambda(x,z) = \epsilon [Q,\lambda(x,z)] = \epsilon c(x,z) , \\ && \delta_{Q} \pi(x,z) = \epsilon [Q,\pi(x,z)] = 0 , \\ && \delta_{Q} \bar{c}(x,z) = \epsilon [Q,\bar{c}(x,z)] = - \epsilon \pi(x,z) , \\ && \delta_{Q} c(x,z) = \epsilon [Q,c(x,z)] = 0 , \eqa
where the first BRST charge $Q$ is
\bqa && Q = c(x,z) \frac{\delta }{\delta \lambda(x,z)} - \pi(x,z) \frac{\delta }{\delta \bar{c}(x,z)} . \eqa 
Here, the infinitesimal parameter $\epsilon$ is fermionic. 
Then, the bulk effective Lagrangian is transformed into
\bqa && \delta_{Q} \Big\{ \pi(x,z) \Big( \partial_{z} \lambda(x,z) - \beta[\lambda(x,z);z] \Big) - \frac{\Gamma}{2} \pi^{2}(x,z) + \bar{c}(x,z) \Big(\partial_{z} - \frac{\partial \beta[\lambda(x,z);z]}{\partial \lambda(x,z)} \Big) c(x,z) + \mathcal{V}_{rg}[\lambda(x,z);z] \Big\} \nn && = \delta_{Q} \mathcal{V}_{rg}[\lambda(x,z);z] = - \epsilon c(x,z) \beta[\lambda(x,z);z] , \eqa
which is not invariant due to the effective potential $\mathcal{V}_{rg}[\lambda(x,z);z]$. 

The second BRST transformation is given by
\bqa && \delta_{\bar{Q}} \lambda(x,z) = \bar{\epsilon} [\bar{Q},\lambda(x,z)] = \bar{\epsilon} \bar{c}(x,z) , \\ && \delta_{\bar{Q}} \pi(x,z) = \bar{\epsilon} [\bar{Q},\pi(x,z)] =  \bar{\epsilon} \frac{2}{\Gamma} \partial_{z} \bar{c}(x,z) , \\ && \delta_{\bar{Q}} \bar{c}(x,z)  = \bar{\epsilon} [\bar{Q},\bar{c}(x,z)] = 0 , \\ && \delta_{\bar{Q}} c(x,z) = \bar{\epsilon} [\bar{Q},c(x,z)] = \bar{\epsilon} \Big( \pi(x,z) - \frac{2}{\Gamma} \partial_{z} \lambda(x,z) \Big) , \eqa
where the second BRST charge $\bar{Q}$ is
\bqa && \bar{Q} = \bar{c}(x,z) \frac{\delta }{\delta \lambda(x,z)} + \frac{2}{\Gamma} [\partial_{z} \bar{c}(x,z)] \frac{\delta }{\delta \pi(x,z)} + \Big( \pi(x,z) - \frac{2}{\Gamma} \partial_{z} \lambda(x,z) \Big) \frac{\delta }{\delta c(x,z)} . \eqa 
As a result, the bulk effective Lagrangian is transformed into
\bqa && \delta_{\bar{Q}} \Big\{ \pi(x,z) \Big( \partial_{z} \lambda(x,z) - \beta[\lambda(x,z);z] \Big) - \frac{\Gamma}{2} \pi^{2}(x,z) + \bar{c}(x,z) \Big(\partial_{z} - \frac{\partial \beta[\lambda(x,z);z]}{\partial \lambda(x,z)} \Big) c(x,z) + \mathcal{V}_{rg}[\lambda(x,z);z] \Big\} \nn && = \bar{\epsilon} \frac{d}{d z} \Big\{ \frac{2}{\Gamma} \bar{c}(x,z) \Big( \partial_{z} \lambda(x,z) - \beta[\lambda(x,z);z] \Big) - \bar{c}(x,z) \pi(x,z) \Big\} - \bar{\epsilon} \bar{c}(x,z) \beta[\lambda(x,z);z] , \eqa
where $\delta_{\bar{Q}} \mathcal{V}_{rg}[\lambda(x,z);z] = - \bar{\epsilon} \bar{c}(x,z) \beta[\lambda(x,z);z]$. The total derivative term does not affect the corresponding Ward identity to be discussed below.  

The third BRST transformation is given by
\bqa && \delta_{D} \lambda(x,z) = \varepsilon [D,\lambda(x,z)] = \varepsilon \bar{c}(x,z) , \\ && \delta_{D} \pi(x,z) = \varepsilon [D,\pi(x,z)] = 0 , \\ && \delta_{D} \bar{c}(x,z) = \varepsilon [D,\bar{c}(x,z)] = 0 , \\ && \delta_{D} c(x,z) = \varepsilon [D,c(x,z)] = \varepsilon \pi(x,z) , \eqa
where the third BRST charge $D$ is
\bqa && D = \bar{c}(x,z) \frac{\delta }{\delta \lambda(x,z)} + \pi(x,z) \frac{\delta }{\delta c(x,z)} . \eqa 
Accordingly, the bulk Lagrangian transforms as
\bqa && \delta_{D} \Big\{ \pi(x,z) \Big( \partial_{z} \lambda(x,z) - \beta[\lambda(x,z);z] \Big) - \frac{\Gamma}{2} \pi^{2}(x,z) + \bar{c}(x,z) \Big(\partial_{z} - \frac{\partial \beta[\lambda(x,z);z]}{\partial \lambda(x,z)} \Big) c(x,z) + \mathcal{V}_{rg}[\lambda(x,z);z] \Big\} \nn && = 2 \varepsilon \pi(x,z) \partial_{z} \bar{c}(x,z) - \varepsilon \partial_{z} \Big( \bar{c}(x,z) \pi(x,z) \Big) - \varepsilon \bar{c}(x,z) \beta[\lambda(x,z);z] , \eqa
where $\delta_{D} \mathcal{V}_{rg}[\lambda(x,z);z] = - \varepsilon \bar{c}(x,z) \beta[\lambda(x,z);z]$. We point out an additional noninvariant term $2 \varepsilon \pi(x,z) \partial_{z} \bar{c}(x,z)$.

The last BRST transformation is given by
\bqa && \delta_{\bar{D}} \lambda(x,z) = \bar{\varepsilon} [\bar{D},\lambda(x,z)] = \bar{\varepsilon} c(x,z) , \\ && \delta_{\bar{D}} \pi(x,z) = \bar{\varepsilon} [\bar{D},\pi(x,z)] =  \bar{\varepsilon} \frac{2}{\Gamma} \partial_{z} c(x,z) , \\ && \delta_{\bar{D}} \bar{c}(x,z)  = \bar{\varepsilon} [\bar{D},\bar{c}(x,z)] = - \bar{\varepsilon} \Big( \pi(x,z) - \frac{2}{\Gamma} \partial_{z} \lambda(x,z) \Big) , \\ && \delta_{\bar{D}} c(x,z) = \bar{\varepsilon} [\bar{D},c(x,z)] = 0 , \eqa
where the last BRST charge $\bar{D}$ is
\bqa && \bar{D} = c(x,z) \frac{\delta }{\delta \lambda(x,z)} + \frac{2}{\Gamma} [\partial_{z} c(x,z)] \frac{\delta }{\delta \pi(x,z)} - \Big( \pi(x,z) - \frac{2}{\Gamma} \partial_{z} \lambda(x,z) \Big) \frac{\delta }{\delta \bar{c}(x,z)} . \eqa  
The bulk effective Lagrangian transforms as
\bqa && \delta_{\bar{D}} \Big\{ \pi(x,z) \Big( \partial_{z} \lambda(x,z) - \beta[\lambda(x,z);z] \Big) - \frac{\Gamma}{2} \pi^{2}(x,z) + \bar{c}(x,z) \Big(\partial_{z} - \frac{\partial \beta[\lambda(x,z);z]}{\partial \lambda(x,z)} \Big) c(x,z) + \mathcal{V}_{rg}[\lambda(x,z);z] \Big\} \nn && = - 2 \bar{\varepsilon} \pi(x,z) [\partial_{z} c(x,z)] + \bar{\varepsilon} \frac{4}{\Gamma} [\partial_{z} c(x,z)] [\partial_{z} \lambda(x,z)] - \bar{\varepsilon} \frac{2}{\Gamma} \partial_{z} \Big( c(x,z) \beta[\lambda_{a}(x,z);z] \Big) - \bar{\varepsilon} c(x,z) \beta[\lambda(x,z);z] , \eqa
where $\delta_{\bar{D}} \mathcal{V}_{rg}[\lambda(x,z);z] = - \bar{\varepsilon} c(x,z) \beta[\lambda(x,z);z]$. We point out an additional noninvariant term $- 2 \bar{\varepsilon} \pi(x,z) [\partial_{z} c(x,z)] + \bar{\varepsilon} \frac{4}{\Gamma} [\partial_{z} c(x,z)] [\partial_{z} \lambda(x,z)]$.

Before discussing the Ward identities for correlation functions 
of the coupling field, 
we check out anti-commutators between BRST charges. 
The anti-commutator for the first two BRST charges is given by
\begin{small}
\bqa && [Q, \bar{Q}]_{+} = Q \bar{Q} + \bar{Q} Q = - \frac{2}{\Gamma} \Big( \pi(x,z) \partial_{z} \frac{\delta }{\delta \pi(x,z)} + \partial_{z} \lambda(x,z) \frac{\delta }{\delta \lambda(x,z)} + c(x,z) \partial_{z} \frac{\delta }{\delta c(x,z)} + [ \partial_{z} \bar{c}(x,z)] \frac{\delta }{\delta \bar{c}(x,z)} \Big) , \eqa
\end{small}
and that of the last two BRST ones is given by
\begin{small}
\bqa && [D, \bar{D}]_{+} = D \bar{D} + \bar{D} D = \frac{2}{\Gamma} \Big( \pi(x,z) \partial_{z} \frac{\delta }{\delta \pi(x,z)} + \partial_{z} \lambda(x,z) \frac{\delta }{\delta \lambda(x,z)} + [\partial_{z} c(x,z)] \frac{\delta }{\delta c(x,z)} + \bar{c}(x,z) \partial_{z} \frac{\delta }{\delta \bar{c}(x,z)} \Big) . \eqa
\end{small} 
In Appendix B, we revisit these two anti-commutation relations between BRST charges in the superspace formulation.

\subsection{Generalized fluctuation-dissipation theorems for the RG flows of correlation functions of the coupling functions} 

To derive the Ward identities from these BRST transformations, 
we consider an action for sources as follows \cite{MSR_Formulation_SUSY_iv,QFT_textbook}
\bqa && \mathcal{S}_{Source} = N \int_{\Lambda_{uv}}^{z_{f}} d z \int d^{D} x \Big( \bar{T}(x,z) \lambda(x,z) + \pi(x,z) T(x,z) + \bar{G}(x,z) c(x,z) + \bar{c}(x,z) G(x,z) \Big) \nn && = N \int_{\Lambda_{uv}}^{z_{f}} d z \int d^{D} x \Big( \bar{T}(x,z) \frac{\partial}{\partial \bar{T}(x,z)} + T(x,z) \frac{\partial}{\partial T(x,z)} + \bar{G}(x,z) \frac{\partial}{\partial \bar{G}(x,z)} - \frac{\partial}{\partial G(x,z)} G(x,z) \Big) \nn && = N \int_{\Lambda_{uv}}^{z_{f}} d z \int d^{D} x \Big( \frac{\partial}{\partial \lambda(x,z)} \lambda(x,z) + \frac{\partial}{\partial \pi(x,z)} \pi(x,z) - \frac{\partial}{\partial c(x,z)} c(x,z) + \bar{c}(x,z) \frac{\partial}{\partial \bar{c}(x,z)} \Big) . \eqa 
Here, $\bar{T}(x,z)$ ($T(x,z)$) is the bosonic source field for $\lambda(x,z)$ ($\pi(x,z)$), 
and $\bar{G}(x,z)$ ($G(x,z)$) is the fermionic source field for $c(x,z)$ ($\bar{c}(x,z)$). 
Accordingly, the four BRST charges are represented as follows 
\bqa && Q = c(x,z) \frac{\delta }{\delta \lambda(x,z)} - \pi(x,z) \frac{\delta }{\delta \bar{c}(x,z)} = \bar{T}(x,z) \frac{\partial}{\partial \bar{G}(x,z)} - G(x,z) \frac{\partial}{\partial T(x,z)} , \\ && \bar{Q} = \bar{c}(x,z) \frac{\delta }{\delta \lambda(x,z)} + \frac{2}{\Gamma} [\partial_{z} \bar{c}(x,z)] \frac{\delta }{\delta \pi(x,z)} + \Big( \pi(x,z) - \frac{2}{\Gamma} \partial_{z} \lambda(x,z) \Big) \frac{\delta }{\delta c(x,z)} \nn && = - \bar{T}(x,z) \frac{\partial}{\partial G(x,z)} - \frac{2}{\Gamma} T(x,z) \Big(\partial_{z} \frac{\partial}{\partial G(x,z)}\Big) - \bar{G}(x,z) \Big( \frac{\partial}{\partial T(x,z)} - \frac{2}{\Gamma} \partial_{z} \frac{\partial}{\partial \bar{T}(x,z)} \Big) , \eqa 
and
\bqa && D = \bar{c}(x,z) \frac{\delta }{\delta \lambda(x,z)} + \pi(x,z) \frac{\delta }{\delta c(x,z)} = - \bar{T}(x,z) \frac{\partial}{\partial G(x,z)} - \bar{G}(x,z) \frac{\partial}{\partial T(x,z)} , \\ && \bar{D} = c(x,z) \frac{\delta }{\delta \lambda(x,z)} + \frac{2}{\Gamma} [\partial_{z} c(x,z)] \frac{\delta }{\delta \pi(x,z)} - \Big( \pi(x,z) - \frac{2}{\Gamma} \partial_{z} \lambda(x,z) \Big) \frac{\delta }{\delta \bar{c}(x,z)} \nn && = \bar{T}(x,z) \frac{\partial}{\partial \bar{G}(x,z)} + \frac{2}{\Gamma} T(x,z) \Big(\partial_{z} \frac{\partial}{\partial \bar{G}(x,z)}\Big) - G(x,z) \Big( \frac{\partial}{\partial T(x,z)} - \frac{2}{\Gamma} \partial_{z} \frac{\partial}{\partial \bar{T}(x,z)} \Big) , \eqa
respectively.

Taking the first two BRST transformations to the partition function, 
we find 
\begin{small}
\bqa && \int_{\Lambda_{uv}}^{z_{f}} d z \int d^{D} x \Big(\bar{T}(x,z) \frac{\partial}{\partial \bar{G}(x,z)} - G(x,z) \frac{\partial}{\partial T(x,z)}\Big) Z(z_{f}) = \frac{1}{2} \int_{\Lambda_{uv}}^{z_{f}} d z \int d^{D} x \beta\Big(\frac{\partial}{\partial \bar{T}(x,z)}; z\Big) \frac{\partial}{\partial \bar{G}(x,z)} Z(z_{f}) , \label{BRST_I_Z} \\ \nn && \int_{\Lambda_{uv}}^{z_{f}} d z \int d^{D} x \Big\{ \bar{T}(x,z) \frac{\partial}{\partial G(x,z)} + \frac{2}{\Gamma} T(x,z) \Big(\partial_{z} \frac{\partial}{\partial G(x,z)}\Big) + \bar{G}(x,z) \Big( \frac{\partial}{\partial T(x,z)} - \frac{2}{\Gamma} \partial_{z} \frac{\partial}{\partial \bar{T}(x,z)} \Big) \Big\} Z(z_{f}) \nn && = \frac{1}{2} \int_{\Lambda_{uv}}^{z_{f}} d z \int d^{D} x \Big[ - \partial_{z} \Big\{ \frac{2}{\Gamma} \Big( \partial_{z} \frac{\partial}{\partial \bar{T}(x,z)} - \beta\Big(\frac{\partial}{\partial \bar{T}(x,z)}; z\Big) \Big) \frac{\partial}{\partial G(x,z)} - \frac{\partial}{\partial T(x,z)} \frac{\partial}{\partial G(x,z)} \Big\} + \beta\Big(\frac{\partial}{\partial \bar{T}(x,z)}; z\Big) \frac{\partial}{\partial G(x,z)} \Big] Z(z_{f}) . \nn \label{BRST_II_Z} \eqa
\end{small} 
Considering the second two BRST transformations to the partition function, we obtain 
\begin{small}
\bqa && \int_{\Lambda_{uv}}^{z_{f}} d z \int d^{D} x \Big( \bar{T}(x,z) \frac{\partial}{\partial G(x,z)} + \bar{G}(x,z) \frac{\partial}{\partial T(x,z)}\Big) Z(z_{f}) \nn && = \frac{1}{2} \int_{\Lambda_{uv}}^{z_{f}} d z \int d^{D} x \Big[ \partial_{z} \Big( \frac{\partial}{\partial T(x,z)} \frac{\partial}{\partial G(x,z)} \Big) - 2 \frac{\partial}{\partial T(x,z)} \partial_{z} \frac{\partial}{\partial G(x,z)} + \beta\Big(\frac{\partial}{\partial \bar{T}(x,z)}; z\Big) \frac{\partial}{\partial G(x,z)} \Big] Z(z_{f}) , \label{BRST_III_Z} \\ \nn && \int_{\Lambda_{uv}}^{z_{f}} d z \int d^{D} x \Big\{ \bar{T}(x,z) \frac{\partial}{\partial \bar{G}(x,z)} + \frac{2}{\Gamma} T(x,z) \Big(\partial_{z} \frac{\partial}{\partial \bar{G}(x,z)}\Big) - G(x,z) \Big( \frac{\partial}{\partial T(x,z)} - \frac{2}{\Gamma} \partial_{z} \frac{\partial}{\partial \bar{T}(x,z)} \Big) \Big\} Z(z_{f}) \nn && = \frac{1}{2} \int_{\Lambda_{uv}}^{z_{f}} d z \int d^{D} x \Big[ \frac{2}{\Gamma} \partial_{z} \Big\{ \beta\Big(\frac{\partial}{\partial \bar{T}(x,z)}; z\Big) \frac{\partial}{\partial \bar{G}(x,z)} \Big\} + 2 \frac{\partial}{\partial T(x,z)} \partial_{z} \frac{\partial}{\partial \bar{G}(x,z)} - \frac{4}{\Gamma} \Big(\partial_{z} \frac{\partial}{\partial \bar{T}(x,z)}\Big) \partial_{z} \frac{\partial}{\partial \bar{G}(x,z)} \nn && + \beta\Big(\frac{\partial}{\partial \bar{T}(x,z)}; z\Big) \frac{\partial}{\partial \bar{G}(x,z)} \Big] Z(z_{f}) . \label{BRST_IV_Z} \eqa
\end{small} 

Eqs.\ (\ref{BRST_I_Z}), (\ref{BRST_II_Z}), (\ref{BRST_III_Z}), and (\ref{BRST_IV_Z}) 
are one of the main results of this study. 
Based on these four types of equations, 
one can derive various Ward identities for correlation functions of the coupling field. 
Here, we demonstrate some of them.

Applying $\frac{\partial}{\partial G(x',z')} \frac{\partial}{\partial \bar{T}(x'',z'')}$ 
to Eq.\ (\ref{BRST_I_Z}) and $\frac{\partial}{\partial \bar{G}(x',z')} \frac{\partial}{\partial \bar{T}(x'',z'')}$ 
to Eq.\ (\ref{BRST_II_Z}), respectively, we obtain 
\bqa && \Big(\frac{\partial}{\partial G(x',z')} \frac{\partial}{\partial \bar{G}(x'',z'')} - \frac{\partial}{\partial \bar{T}(x'',z'')} \frac{\partial}{\partial T(x',z')} \Big) Z(z_{f}) \nn && = \frac{1}{2} \frac{\partial}{\partial G(x',z')} \frac{\partial}{\partial \bar{T}(x'',z'')} \int_{\Lambda_{uv}}^{z_{f}} d z \int d^{D} x \beta\Big(\frac{\partial}{\partial \bar{T}(x,z)}; z\Big) \frac{\partial}{\partial \bar{G}(x,z)} Z(z_{f}) , \label{Ward_Identity_I_Diff} \\ && \Big( \frac{\partial}{\partial \bar{G}(x',z')} \frac{\partial}{\partial G(x'',z'')} + \frac{\partial}{\partial \bar{T}(x'',z'')} \frac{\partial}{\partial T(x',z')} - \frac{2}{\Gamma} \frac{\partial}{\partial \bar{T}(x'',z'')} \partial_{z'} \frac{\partial}{\partial \bar{T}(x',z')} \Big) Z(z_{f}) \nn && = \frac{1}{2} \frac{\partial}{\partial \bar{G}(x',z')} \frac{\partial}{\partial \bar{T}(x'',z'')} \int_{\Lambda_{uv}}^{z_{f}} d z \int d^{D} x \beta\Big(\frac{\partial}{\partial \bar{T}(x,z)}; z\Big)  \frac{\partial}{\partial G(x,z)} Z(z_{f}) , \label{Ward_Identity_II_Diff} \eqa
where all other source fields were set to zero. These two equations
%
%Eqs. (\ref{Ward_Identity_I_Diff}) and (\ref{Ward_Identity_II_Diff}) 
%
lead to
\begin{small}
\bqa && \Big\langle \bar{c}(x',z') c(x,z) \Big\rangle + \Big\langle \lambda(x,z) \pi(x',z') \Big\rangle = - \frac{1}{2} \Big\langle \bar{c}(x',z') \lambda(x,z) \int_{\Lambda_{uv}}^{z_{f}} d w \int d^{D} y \beta[\lambda(y,w);w] c(y,w) \Big\rangle , \label{Ward_Identity_Corr_I} \\ && \Big\langle c(x',z') \bar{c}(x,z) \Big\rangle - \Big\langle \lambda(x,z) \pi(x',z') \Big\rangle + \frac{2}{\Gamma} \Big\langle \lambda(x,z) \partial_{z'} \lambda(x',z') \Big\rangle = \frac{1}{2} \Big\langle c(x',z') \lambda(x,z) \int_{\Lambda_{uv}}^{z_{f}} d w \int d^{D} y \beta[\lambda(y,w);w] \bar{c}(y,w) \Big\rangle , \nn \label{Ward_Identity_Corr_II}  \eqa
\end{small} 
respectively. 

Considering the ghost Green's function 
$\Big(\partial_{z} - \frac{\partial \beta[\lambda(x,z);z]}{\partial \lambda(x,z)} \Big) \langle c(x,z) \bar{c}(x',z') \rangle = - \delta^{(D)}(x-x') \delta(z-z')$, 
we obtain
\begin{small}
\bqa && \Big\langle \lambda(x',z') \pi(x,z) \Big\rangle - \Big\langle \lambda(x,z) \pi(x',z') \Big\rangle = - \frac{2}{\Gamma} \Big\langle \lambda(x,z) \partial_{z'} \lambda(x',z') \Big\rangle - \Big\langle \lambda(x',z') \Big(\partial_{z} - \frac{\partial \beta[\lambda(x,z);z]}{\partial \lambda(x,z)} \Big)^{-1} \beta[\lambda(x,z);z] \Big\rangle \label{Fluctuation_Dissipation} \eqa
\end{small} 
from Eqs.\ (\ref{Ward_Identity_Corr_I}) and (\ref{Ward_Identity_Corr_II}). 
If we consider a fixed point defined by $\beta[\lambda(x,z);z] = 0$, we obtain
\bqa && \Big\langle \lambda(x',z') \pi(x,z) \Big\rangle - \Big\langle \lambda(x,z) \pi(x',z') \Big\rangle = - \frac{2}{\Gamma} \Big\langle \lambda(x,z) \partial_{z'} \lambda(x',z') \Big\rangle . \nonumber \eqa
This is essentially the same as the fluctuation-dissipation theorem 
of the Langevin dynamics in equilibrium. 
Away from the fixed point, 
there is an RG flow given by the RG $\beta$-function, 
which plays the role of the nonequilibrium work in the dynamics, 
reflected in the last term of Eq.\ (\ref{Fluctuation_Dissipation}). 

%
%Similarly, we can find a Jarzynski-like identity. Applying $\frac{\partial}{\partial \bar{G}(x',z')}$ to Eq. (\ref{BRST_II_Z}), where all other source fields are set to be zero, we obtain
%\begin{small}
%\bqa &&  \frac{\partial}{\partial \bar{G}(x',z')} \Big( \frac{\partial}{\partial T(x',z')} - \frac{2}{\Gamma} \partial_{z'} \frac{\partial}{\partial \bar{T}(x',z')} \Big) Z(z_{f}) = \frac{1}{2} \frac{\partial}{\partial %\bar{G}(x',z')} \int_{\Lambda_{uv}}^{z_{f}} d z \int d^{D} x \beta\Big(\frac{\partial}{\partial \bar{T}(x,z)}; z\Big)  \frac{\partial}{\partial G(x,z)} Z(z_{f}) . \eqa
%\end{small}
%This equation is easily translated into
%\begin{small}
%\bqa && \Big\langle c(x,z) \Big( \pi(x,z) - \frac{2}{\Gamma} \partial_{z} \lambda(x,z) \Big) \Big\rangle = - \frac{1}{2} \Big\langle c(x,z) \int_{\Lambda_{uv}}^{z_{f}} d w \int d^{D} y %\beta[\lambda(y,w);w] \bar{c}(y,w) \Big\rangle . \eqa
%\end{small}
%We observe that this equation is an expansion of the following exponential form in the first order,
%\begin{small}
%\bqa && \Big\langle c(x,z) \Big( \pi(x,z) - \frac{2}{\Gamma} \partial_{z} \lambda(x,z) \Big) \exp\Big\{\frac{1}{2} \int_{\Lambda_{uv}}^{z_{f}} d w \int d^{D} y \beta[\lambda(y,w);w] \bar{c}(y,w)\Big\} \Big\rangle = 0 . \eqa
%\end{small} 
%It has been shown that this expression is identical to the Jarzynski identity in nonequilibrium thermodynamics.
%

\subsection{RG flow of an on-shell effective action: Hamilton-Jacobi equation} 

To discuss the RG flow of an IR effective action, 
we take the large $N$ limit and obtain equations of motion with boundary conditions. 
We recall the holographic dual field theory,
\begin{small}
\bqa && Z(z_{f}) = \int D \psi_{\sigma}(x,z_{f}) D \lambda(x,z) D \pi(x,z) D \bar{c}(x,z) D c(x,z) \exp\Big[ - \int d^{D} x \Big( \mathcal{L}[\psi_{\sigma}(x,z_{f});\lambda(x,z_{f});z_{f}] + \frac{N}{2 \Gamma}[\lambda(x,\Lambda_{uv}) - \bar{\lambda}(\Lambda_{uv})]^{2} \Big) \nn && - N \int_{\Lambda_{uv}}^{z_{f}} d z \int d^{D} x \Big\{ \pi(x,z) \Big( \partial_{z} \lambda(x,z) - \beta[\lambda(x,z);z] \Big) - \frac{\Gamma}{2} \pi^{2}(x,z) + \bar{c}(x,z) \Big(\partial_{z} - \frac{\partial \beta[\lambda(x,z);z]}{\partial \lambda(x,z)} \Big) c(x,z) + \mathcal{V}_{rg}[\lambda(x,z);z] \Big\} \Big] . \nonumber \eqa  
\end{small}
Taking the large $N$ limit and performing variations with respect to $\pi(x,z)$ and $\lambda(x,z)$, 
we obtain the Hamiltonian equation of motion as follows
\bqa && \pi(x,z) = \frac{1}{\Gamma} \Big( \partial_{z} \lambda(x,z) - \beta[\lambda(x,z);z] \Big) \label{Canonical_Momentum} , \\ && \partial_{z} \pi(x,z) = - \pi(x,z) \frac{\partial \beta[\lambda(x,z);z]}{\partial \lambda(x,z)} - \Big(\partial_{z} - \frac{\partial \beta[\lambda(x,z);z]}{\partial \lambda(x,z)} \Big)^{-1} \frac{\partial^{2} \beta[\lambda(x,z);z]}{\partial \lambda^{2}(x,z)} - \beta[\lambda(x,z);z] . \eqa
We recall 
\bqa && \beta[\lambda(x,z);z] = - \frac{\partial \mathcal{V}_{rg}[\lambda(x,z);z]}{\partial \lambda(x,z)} . \nonumber \eqa

Just for completeness, we write down the Lagrangian formulation for the partition function,
\begin{small}
\bqa && Z(z_{f}) = \int D \psi_{\sigma}(x,z_{f}) D \lambda(x,z) D \bar{c}(x,z) D c(x,z) \exp\Big[ - \int d^{D} x \Big( \mathcal{L}[\psi_{\sigma}(x,z_{f});\lambda(x,z_{f});z_{f}] + \frac{N}{2 \Gamma}[\lambda(x,\Lambda_{uv}) - \bar{\lambda}(\Lambda_{uv})]^{2} \Big) \nn && - N \int_{\Lambda_{uv}}^{z_{f}} d z \int d^{D} x \Big\{ \frac{1}{2 \Gamma} \Big( \partial_{z} \lambda(x,z) - \beta[\lambda(x,z);z] \Big)^{2} + \bar{c}(x,z) \Big(\partial_{z} - \frac{\partial \beta[\lambda(x,z);z]}{\partial \lambda(x,z)} \Big) c(x,z) + \mathcal{V}_{rg}[\lambda(x,z);z] \Big\} \Big] , \eqa  
\end{small}
and obtain the corresponding Lagrange equation of motion,
\bqa && \frac{1}{\Gamma} \partial_{z}^{2} \lambda(x,z) = - \beta[\lambda(x,z);z] + \frac{1}{\Gamma} \beta[\lambda(x,z);z] \frac{\partial \beta[\lambda(x,z);z]}{\partial \lambda(x,z)} - \Big(\partial_{z} - \frac{\partial \beta[\lambda(x,z);z]}{\partial \lambda(x,z)} \Big)^{-1} \frac{\partial^{2} \beta[\lambda(x,z);z]}{\partial \lambda^{2}(x,z)} . \eqa

To obtain boundary conditions, we consider an effective boundary action as
\bqa && \mathcal{S}_{eff}(z_{f}) = N \int d^{D} x \Big( \mathcal{V}_{rg}[\lambda(x,z_{f});z_{f}] + \pi(x,z_{f}) \lambda(x,z_{f}) + \bar{c}(x,z_{f}) c(x,z_{f}) \nn && + \frac{N}{2 \Gamma}[\lambda(x,\Lambda_{uv}) - \bar{\lambda}(\Lambda_{uv})]^{2} - \pi(x,\Lambda_{uv}) \lambda(x,\Lambda_{uv}) - \bar{c}(x,\Lambda_{uv}) c(x,\Lambda_{uv}) \Big) . \label{IR_Boundary_Effective_Action} \eqa
Here, the boundary effective potential $\mathcal{V}_{rg}[\lambda(x,z_{f});z_{f}]$ comes from
\bqa && \mathcal{V}_{rg}[\lambda(x,z_{f});z_{f}] = - \frac{1}{N} \ln \int_{\Lambda(z_{f})} D \psi_{\sigma}(x;z_{f}) \exp\Big\{ - \int d^{D} x \mathcal{L}[\psi_{\sigma}(x,z_{f});\lambda(x,z_{f});z_{f}] \Big\} , \label{IR_Effective_Potential} \eqa
and $\pi(x,z_{f}) \lambda(x,z_{f}) + \bar{c}(x,z_{f}) c(x,z_{f}) - \pi(x,\Lambda_{uv}) \lambda(x,\Lambda_{uv}) - \bar{c}(x,\Lambda_{uv}) c(x,\Lambda_{uv})$ results from the bulk action $\pi(x,z) \partial_{z} \lambda(x,z) + \bar{c}(x,z) \partial_{z} c(x,z) $ by integration by parts. Taking variations of this effective boundary action with respect to $\lambda(x,z_{f})$, $c(x,z_{f})$ and $\lambda(x,\Lambda_{uv})$, $c(x,\Lambda_{uv})$, we find the IR boundary conditions 
\bqa && \pi(x,z_{f}) = \beta[\lambda(x,z_{f});z_{f}] , ~~~~~ \bar{c}(x,z_{f}) = 0 , \eqa
and UV ones
\bqa && \pi(x,\Lambda_{uv}) = \frac{N}{\Gamma}[\lambda(x,\Lambda_{uv}) - \bar{\lambda}(\Lambda_{uv})] , ~~~~~ \bar{c}(x,\Lambda_{uv}) = 0 . \eqa 
Combined with the canonical momentum Eq.\ (\ref{Canonical_Momentum}), 
we obtain both UV and IR boundary conditions for $\lambda(x,z)$, 
which support the second order Lagrange equation of motion for $\lambda(x,z)$.

We emphasize again that the partition function is invariant under the RG transformation, 
formulated as
\bqa && - \frac{d }{d z_{f}} \ln Z(z_{f}) = 0 . \eqa 
This equation gives rise to
%
%\begin{small}
%\bqa && - \Big( \frac{\Gamma}{2} + 1 \Big) \Big(\frac{\partial \mathcal{V}_{rg}[\lambda(x,z_{f});z_{f}]}{\partial \lambda(x,z_{f})}\Big)^{2} + \bar{c}(x,z_{f}) \Big(\partial_{z_{f}} + \frac{\partial^{2} \mathcal{V}_{rg}[\lambda(x,z_{f});z_{f}]}{\partial \lambda^{2}(x,z_{f})} \Big) c(x,z_{f}) + \mathcal{V}_{rg}[\lambda(x,z_{f});z_{f}] + \partial_{z_{f}} \mathcal{V}_{rg}[\lambda(x,z_{f});z_{f}] = 0 , \nn \eqa  
%\end{small}
%
\bqa && - \Big( \frac{\Gamma}{2} + 1 \Big) \Big(\frac{\partial \mathcal{V}_{rg}[\lambda(x,z_{f});z_{f}]}{\partial \lambda(x,z_{f})}\Big)^{2} + \mathcal{V}_{rg}[\lambda(x,z_{f});z_{f}] + \partial_{z_{f}} \mathcal{V}_{rg}[\lambda(x,z_{f});z_{f}] = 0 , \eqa  
where the IR boundary condition $\pi(x,z_{f}) = \beta[\lambda(x,z_{f});z_{f}]$ has been used. 
This is nothing but the Hamilton-Jacobi equation to determine the IR renormalized 
effective potential 
$\mathcal{V}_{rg}[\lambda(x,z_{f});z_{f}]$ \cite{Nonperturbative_Wilson_RG, Einstein_Klein_Gordon_RG_Kim}. 
One may regard this Hamilton-Jacobi equation as a signature to guarantee self-consistency 
of the present framework \cite{Nonperturbative_Wilson_RG, Einstein_Klein_Gordon_RG_Kim}, 
where the IR effective potential is given by Eq.\ (\ref{IR_Effective_Potential}).

\subsection{``Entropy production" in the RG flow}

To investigate the monotonicity or ``irreversibility" of the RG flow, 
we discuss ``entropy production" in the RG flow, 
following the procedure for the overdamped Langevin system in Ref.\ \cite{Entropy_Production}. 
We recall the effective partition function to manifest the RG flow 
as an effective bulk action with an extradimension,
\begin{small}
\bqa && Z(z_{f}) = \int_{\xi_{uv}}^{\xi_{ir}} D \xi(x,z) \exp\Big\{ - N \int_{\Lambda_{uv}}^{z_{f}} d z \int d^{D} x \frac{1}{2 \Gamma} \xi^{2}(x,z) \Big\} \int_{\lambda_{uv}}^{\lambda_{ir}} D \lambda(x,z) D \pi(x,z) D \bar{c}(x,z) D c(x,z) \nn && \exp\Big[ - N \int_{\Lambda_{uv}}^{z_{f}} d z \int d^{D} x \Big\{ \pi(x,z) \Big( \partial_{z} \lambda(x,z) - \beta[\lambda(x,z);z] - \xi(x,z) \Big) + \bar{c}(x,z) \Big(\partial_{z} - \frac{\partial \beta[\lambda(x,z);z]}{\partial \lambda(x,z)} \Big) c(x,z) + \mathcal{V}_{rg}[\lambda(x,z);z] \Big\} \Big] . \nn \eqa  
\end{small}
Here, random noise fluctuations were explicitly introduced 
by the Hubbard-Stratonovich transformation for the bulk canonical momentum. 
The IR boundary conditions were assumed, 
where both $\lambda_{ir}$ and $\xi_{ir}$ 
can be determined by the IR boundary conditions of $\lambda(x,z_{f})$ and $\pi(x,z_{f})$. 
Accordingly, the ``Hamiltonian equation of motion" is given by
\bqa && \partial_{z} \lambda(x,z) = \beta[\lambda(x,z);z] + \xi(x,z) , \label{Eq_Langevin_I} \\ && \partial_{z} \pi(x,z) = - \beta[\lambda(x,z);z] - \pi(x,z) \frac{\partial \beta[\lambda(x,z);z]}{\partial \lambda(x,z)} - \Big(\partial_{z} - \frac{\partial \beta[\lambda(x,z);z]}{\partial \lambda(x,z)} \Big)^{-1} \frac{\partial^{2} \beta[\lambda(x,z);z]}{\partial \lambda^{2}(x,z)} . \label{Eq_Langevin_II} \eqa
The first equation corresponds to the overdamped Langevin equation, 
where the coupling field $\lambda(x,z)$ and the RG scale $z$ 
may be identified with the position of a particle and time.

The ``probability distribution" function for the coupling field is defined as follows
\begin{small}
\begin{align}
\rho(\lambda,z) 
&= \langle \delta(\lambda-\lambda(x,z)) \rangle 
\nonumber \\
&= \mathcal{N} \int D \xi(x,z') \exp\Big\{ - N \int_{\Lambda_{uv}}^{z} d z' \int d^{D} x \Big( \frac{1}{2 \Gamma} \xi^{2}(x,z') + \mathcal{V}_{rg}[\lambda(x,z');z'] \Big) \Big\} \delta(\lambda-\lambda(x,z)) , 
\end{align}
\end{small}where $\mathcal{N}$ is a normalization constant to be specified below. 
We emphasize that there appears a correction in the RG flow, 
given by $\mathcal{V}_{rg}[\lambda(x,z');z']$. 
Then, the path integral expression of this probability distribution function is given by
\begin{small}
\begin{align}
  \rho(\lambda,z) 
  &= \frac{1}{Z(z_{f})} 
  \int_{\lambda_{uv}}^{\lambda} D \lambda(x,z') D \pi(x,z') 
  D \bar{c}(x,z') D c(x,z') 
  \nonumber \\
  &\quad 
  \times 
  \int D \xi(x,z') \exp\Big\{ - N \int_{\Lambda_{uv}}^{z} d z' \int d^{D} x \Big( \frac{1}{2 \Gamma} \xi^{2}(x,z') + \mathcal{V}_{rg}[\lambda(x,z');z'] \Big) \Big\} 
  \nonumber \\
  & \quad 
  \times \exp\Big[ - N \int_{\Lambda_{uv}}^{z} d z' \int d^{D} x 
  \Big\{ \pi(x,z') \Big( \partial_{z'} \lambda(x,z') 
  - \beta[\lambda(x,z');z'] - \xi(x,z') \Big) + \bar{c}(x,z') 
  \Big(\partial_{z'} - \frac{\partial \beta[\lambda(x,z');z']}{\partial \lambda(x,z')} \Big) c(x,z') \Big\} \Big] , 
\end{align}
\end{small}
where the normalization constant is given by the partition function introduced above.
%
%\begin{small}
%\bqa && Z(z) = \int D \xi(x,z') \exp\Big\{ - N \int_{\Lambda_{uv}}^{z} d z' \int d^{D} x \frac{1}{2 \Gamma} \xi^{2}(x,z') \Big\} \int D \lambda(x,z') D \pi(x,z') D \bar{c}(x,z') D c(x,z') \nn && \exp\Big[ - %N \int_{\Lambda_{uv}}^{z} d z' \int d^{D} x \Big\{ \pi(x,z') \Big( \partial_{z'} \lambda(x,z') - \beta[\lambda(x,z');z'] - \xi(x,z') \Big) + \bar{c}(x,z') \Big(\partial_{z'} - \frac{\partial %\beta[\lambda(x,z');z']}{\partial \lambda(x,z')} \Big) c(x,z') \nn && + \mathcal{V}_{rg}[\lambda(x,z');z'] \Big\} \Big] . \eqa  
%\end{small}
%
One can check out
\bqa && \mbox{tr} \rho(\lambda,z) = \int_{\lambda_{uv}}^{\lambda_{ir}} d \lambda \rho(\lambda,z) = 1 . \eqa 

%
%\bqa && \dot{\lambda} \partial_{\lambda} \rho = \partial_{\lambda} (\dot{\lambda} \rho) - (\partial_{\lambda} \dot{\lambda}) \rho - N \dot{\lambda} (\partial_{\lambda} V_{rg}) \rho = %\partial_{\lambda} (\dot{\lambda} \rho) + \dot{\lambda} \beta \rho = \partial_{\lambda} [ (\beta + \xi) \rho] + (\beta + \xi) \beta \rho \nn && \rightarrow  \partial_{\lambda} [ (\beta - \Gamma/2 %\partial_{\lambda}) \rho] + (\beta - \Gamma / 2 \beta) \beta \rho \eqa
%

Following the standard procedure to derive the Fokker-Planck equation 
from the Langevin equation, we obtain
\bqa && \Big( \partial_{z} - \mathcal{V}_{rg}(\lambda,z) \Big) \rho(\lambda,z) = - \partial_{\lambda} \Big\{ \Big( \beta(\lambda,z) - \frac{\Gamma}{2} \partial_{\lambda} \Big) \rho(\lambda,z) \Big\} , \label{Fokker_Planck_Eq_RG} \eqa
where the RG effective potential $\mathcal{V}_{rg}(\lambda,z)$ 
serves as the ``time" component of a background gauge field. 
The conserved current is given by
\bqa &&  j(\lambda,z)  = \Big( \beta(\lambda,z) - \frac{\Gamma}{2} \partial_{\lambda} \Big) \rho(\lambda,z) , \eqa
which shares essentially the same structure as that of the overdamped Langevin dynamics, 
discussed before. 
In Appendix A, we show our intuitive derivation for this Fokker-Planck equation.

%
%$\rho(\lambda,z) = \rho_{uv} \exp[- \mathcal{S}_{eff}(\lambda,z)]$
%\bqa && \Big( \partial_{z} - \mathcal{V}_{rg}(\lambda,z) \Big) \exp[- \mathcal{S}_{eff}(\lambda,z)] = - \partial_{\lambda} \Big\{ \Big( \beta(\lambda,z) - \frac{\Gamma}{2} \partial_{\lambda} \Big) \exp[- \mathcal{S}_{eff}(\lambda,z)] \Big\} , \eqa
%\bqa && \Big( - \partial_{z} \mathcal{S}_{eff}(\lambda,z) - \mathcal{V}_{rg}(\lambda,z) \Big) \exp[- \mathcal{S}_{eff}(\lambda,z)] = - \partial_{\lambda} \Big\{ \Big( \beta(\lambda,z) + \frac{\Gamma}{2} \partial_{\lambda} \mathcal{S}_{eff}(\lambda,z) \Big) \exp[- \mathcal{S}_{eff}(\lambda,z)] \Big\} , \eqa
%\bqa && - \partial_{z} \mathcal{S}_{eff}(\lambda,z) - \mathcal{V}_{rg}(\lambda,z) = - \Big( \partial_{\lambda} \beta(\lambda,z) + \frac{\Gamma}{2} \partial_{\lambda}^{2} \mathcal{S}_{eff}(\lambda,z) \Big) + \partial_{\lambda} \mathcal{S}_{eff}(\lambda,z) \Big( \beta(\lambda,z) + \frac{\Gamma}{2} \partial_{\lambda} \mathcal{S}_{eff}(\lambda,z) \Big) , \eqa
%\bqa && - \Big( \frac{\Gamma}{2} + 1 \Big) \Big(\frac{\partial \mathcal{V}_{rg}[\lambda(x,z_{f});z_{f}]}{\partial \lambda(x,z_{f})}\Big)^{2} + \mathcal{V}_{rg}[\lambda(x,z_{f});z_{f}] + \partial_{z_{f}} \mathcal{V}_{rg}[\lambda(x,z_{f});z_{f}] = 0 , \eqa  
%

Following Ref.\ \cite{Entropy_Production}, 
it is natural to introduce the entropy of a system, given by
\bqa && s_{sys}(\lambda,z) = - \ln \rho(\lambda,z) . \eqa
Then, the ensemble average of the system or bulk entropy is
\bqa && S_{sys}(z) = \langle s_{sys}(\lambda,z) \rangle = - \int_{\lambda_{uv}}^{\lambda_{ir}} d \lambda \rho(\lambda,z) \ln \rho(\lambda,z) ,
\eqa
as expected. 

%
%\begin{small}
%\bqa && \Big\langle \partial_{z} H_{eff}(\lambda,z) \Big\rangle 
%
%= N \Big\langle \int d^{D} x \Big\{ \frac{1}{\Gamma} \xi(x,z) \partial_{z} \xi(x,z) - [\partial_{z} \pi(x,z)] \beta[\lambda(x,z);z] - \pi(x,z) \frac{\partial \beta[\lambda(x,z);z]}{\partial \lambda(x,z)} %[\partial_{z} \lambda(x,z)] - \beta[\lambda(x,z);z] [\partial_{z} \lambda(x,z)] \nn && - \Big(\partial_{z} - \frac{\partial \beta[\lambda(x,z);z]}{\partial \lambda(x,z)} \Big)^{-1} \frac{\partial^{2} %\beta[\lambda(x,z);z]}{\partial \lambda^{2}(x,z)} [\partial_{z} \lambda(x,z)] \Big\} \Big\rangle 
%
%= N \Big\langle \int d^{D} x \xi(x,z) \Big(\frac{1}{\Gamma} \partial_{z} \xi(x,z) + \partial_{z} \pi(x,z) \Big) \Big\rangle = N  \int d^{D} x \frac{\Gamma}{2} [\partial_{z} \pi(x,z)]^{2} \eqa
%\end{small}
%

%
%\bqa && \partial_{z} \lambda(x,z) = \beta[\lambda(x,z);z] + \xi(x,z) , \\ && \partial_{z} \pi(x,z) = - \beta[\lambda(x,z);z] - \pi(x,z) \frac{\partial \beta[\lambda(x,z);z]}{\partial \lambda(x,z)} - %\Big(\partial_{z} - \frac{\partial \beta[\lambda(x,z);z]}{\partial \lambda(x,z)} \Big)^{-1} \frac{\partial^{2} \beta[\lambda(x,z);z]}{\partial \lambda^{2}(x,z)} . \eqa
%

The time evolution of the bulk entropy is given by
\bqa && \partial_{z} s_{sys}(z) = - \frac{\partial_{z} \rho(\lambda,z)}{\rho(\lambda,z)} - \frac{\partial_{\lambda} \rho(\lambda,z)}{\rho(\lambda,z)} \partial_{z} \lambda(x,z) . \eqa
Resorting to the Fokker-Planck equation and considering the definition of the conserved current, 
we rewrite the above expression as follows
\bqa && \partial_{z} s_{sys}(z) = \frac{\partial_{\lambda} j(\lambda,z)}{\rho(\lambda,z)} - \mathcal{V}_{rg}(\lambda,z) + \frac{2}{\Gamma} \frac{j(\lambda,z)}{\rho(\lambda,z)} \partial_{z} \lambda(x,z) - \frac{2}{\Gamma} \beta(\lambda,z) \partial_{z} \lambda(x,z) . \eqa
Here, we introduce the time evolution of the ``environment" entropy 
in a similar way as Ref.\ \cite{Entropy_Production},
\bqa && \partial_{z} s_{env}(\lambda,z) = \partial_{z} q(\lambda,z) = \frac{2}{\Gamma} \beta(\lambda,z) \partial_{z} \lambda(x,z) + \mathcal{V}_{rg}(\lambda,z) . \eqa
$\partial_{z} q(\lambda,z)$ 
is the rate of heat dissipation in the medium, 
where we identify the exchanged heat with an increase in entropy of the medium. 
%
%It is quite interesting to observe that the ensemble average of the rate of heat dissipation in the environment is always positive as follows
%\bqa && \langle \partial_{z} s_{env}(\lambda,z) \rangle = \frac{2}{\Gamma} \langle \xi(x,z) \beta(\lambda,z) \rangle + \frac{2}{\Gamma} \langle [\beta(\lambda,z)]^{2} \rangle + \langle \mathcal{V}_{rg}(\lambda,z) \rangle = \Big( \frac{\Gamma}{2} + \frac{2}{\Gamma}  - 2 \Big) \langle [\beta(\lambda,z)]^{2} \rangle \geq 0 , \eqa
%completely consistent with the information loss during the RG flow. Here, we used $\langle \xi(x,z) \beta(\lambda,z) \rangle = - \frac{\Gamma}{2} [\beta(\lambda,z)]^{2}$.
%

Summing over these two contributions, we obtain
\bqa && \partial_{z} s_{tot}(\lambda,z) = \partial_{z} s_{env}(\lambda,z) + \partial_{z} s_{sys}(\lambda,z) = \frac{\partial_{\lambda} j(\lambda,z)}{\rho(\lambda,z)} + \frac{2}{\Gamma} \frac{j(\lambda,z)}{\rho(\lambda,z)} \partial_{z} \lambda(x,z) , \eqa
fully consistent with that of the overdamped Langevin dynamics \cite{Entropy_Production}, 
although there exists a clear modification in the Fokker-Planck equation, Eq.\ (\ref{Fokker_Planck_Eq_RG}). 
As a result, we find the irreversibility of the RG flow, given by the total entropy function,
\bqa && \partial_{z} S_{tot}(z) = \langle \partial_{z} s_{tot}(\lambda,z) \rangle = \int_{\lambda_{uv}}^{\lambda_{ir}} d \lambda \frac{2}{\Gamma} \frac{j^{2}(\lambda,z)}{\rho(\lambda,z)} \geq 0 , \eqa 
where the ensemble average has been taken, and the following current conservation has been used,
\bqa && \Big\langle \frac{\partial_{\lambda} j(\lambda,z)}{\rho(\lambda,z)} \Big\rangle = \int_{\lambda_{uv}}^{\lambda_{ir}} d \lambda \partial_{\lambda} j(\lambda,z) = 0 . \eqa
More explicitly, we have
\bqa && \langle \partial_{z} s_{tot}(\lambda,z) \rangle = \int_{\lambda_{uv}}^{\lambda_{ir}} d \lambda \rho(\lambda,z) \Big\{ \frac{2}{\Gamma} [\beta(\lambda,z)]^{2} + \frac{\Gamma}{2} \Big(  \partial_{\lambda} \ln \rho(\lambda,z)\Big)^{2} - 2 \beta(\lambda,z) \partial_{\lambda} \ln \rho(\lambda,z) \Big\} \geq 0 . \eqa

\section{Summary and discussion}

In this study, we applied the MSR formulation to the RG flow 
and obtained the holographic dual effective field theory 
to manifest the RG flow at the level of an effective bulk action. 
Here, we observed that four types of BRST transformations 
can give some constraints to the RG flow of the coupling field 
although the RG-generated effective potential breaks such BRST symmetries. 
Resorting to the BRST transformations, 
we derived Ward identities for correlation functions of the coupling field. 
In particular, we found that the fluctuation-dissipation theorem 
is modified by the RG $\beta$-function, 
analogous to the nonequilibrium work relation. 
This becomes more transparent in the superspace formulation.

It is natural to apply the present framework to the holographic renormalization 
\cite{Holographic_Duality_V, Holographic_Duality_VI, Holographic_Duality_VII}. 
To consider the holographic renormalization, 
we introduce the Arnowitt-Deser-Misner (ADM) formalism for general relativity \cite{ADM_Hamiltonian_Formulation}, where the coordinate system is given by
\bqa && d s^{2} = \Big( \mathcal{N}^{2}(x,z) + \mathcal{N}_{\mu}(x,z) \mathcal{N}^{\mu}(x,z) \Big) d z^{2} + 2 \mathcal{N}_{\mu}(x,z) d x^{\mu} d z + g_{\mu\nu}(x,z) d x^{\mu} d x^{\nu} . \eqa
Here, $\mathcal{N}(x,z)$ is the lapse function and $\mathcal{N}_{\mu}(x,z)$ is the shift vector. 
We consider the Gaussian normal coordinate system, 
given by gauge fixing of $\mathcal{N}(x,z) = 1$ and $\mathcal{N}_{\mu}(x,z) = 0$. 
Then, the holographic bulk effective action is
\begin{small}
\bqa && F = - \frac{1}{\beta} \ln \int D g_{\mu\nu}(x,z) D \pi^{\mu\nu}(x,z) D \bar{c}^{\mu\nu}(x,z) D c^{\rho\gamma}(x,z) \exp\Big[ - N_{c}^{2} \int_{0}^{z_{f}} d z \int d^{D} x \Big\{ \pi^{\mu\nu}(x,z) \Big( \partial_{z} g_{\mu\nu}(x,z) - \beta_{\mu\nu}[ g_{\mu\nu}(x,z);z] \Big) \nn && - \frac{\kappa}{2} \frac{1}{\sqrt{g(x,z)}} \pi^{\mu\nu}(x,z) \mathcal{G}_{\mu\nu\rho\gamma}(x,z) \pi^{\rho\gamma}(x,z) + \bar{c}^{\mu\nu}(x,z) \Big( \partial_{z} \mathcal{G}_{\mu\nu\rho\gamma}(x,z) - \frac{\partial }{\partial g^{\rho\gamma}(x,z)} \beta_{\mu\nu}[ g_{\mu\nu}(x,z);z] \Big) c^{\rho\gamma}(x,z) \nn && + \frac{1}{2 \kappa} \sqrt{g(x,z)} \Big( R(x,z) - 2 \Lambda \Big) \Big\} \Big] . \eqa
\end{small}
%{\color{red}What is $N_c$??}
Here, $N_{c}$ is the number of color degrees of freedom in dual quantum field theories.
$\pi^{\mu\nu}(x,z)$ is the canonical momentum of the metric tensor $g_{\mu\nu}(x,z)$, 
and $\mathcal{G}_{\mu\nu\rho\gamma}(x,z)$ is de Witt supermetric \cite{DeWitt_Metric},
\bqa && \mathcal{G}_{\mu\nu\rho\gamma}(x,z) \equiv g_{\mu\rho}(x,z) g_{\nu\gamma}(x,z) - \frac{1}{D-1} g_{\mu\nu}(x,z) g_{\rho\gamma}(x,z) . \eqa
The RG $\beta$-function is introduced for the appearance of the RG flow as follows
\bqa && \beta_{\mu\nu}[ g_{\mu\nu}(x,z);z] = - \frac{\partial}{\partial g^{\mu\nu}(x,z)} \Big\{ \frac{1}{2 \kappa} \sqrt{g(x,z)} \Big( R(x,z) - 2 \Lambda \Big) \Big\} \nn && = - \frac{1}{2 \kappa \sqrt{g(x,z)}} \Big(R_{\mu\nu}(x,z) - \frac{1}{2} R(x,z) g_{\mu\nu}(x,z) + \Lambda g_{\mu\nu}(x,z) \Big) , \eqa
which modifies the holographic dual effective field theory 
\cite{Nonperturbative_Wilson_RG, Einstein_Klein_Gordon_RG_Kim}.

Following the strategy of the present study, we propose the following Ward identity
\begin{small}
\bqa && \Big\langle g_{\mu\nu}(x',z') \pi^{\mu\nu}(x,z) \Big\rangle - \Big\langle g_{\mu\nu}(x,z) \pi^{\mu\nu}(x',z') \Big\rangle = - \frac{2}{\kappa} \Big\langle g_{\mu\nu}(x,z) \mathcal{G}^{\mu\nu\rho\gamma}(x',z') \partial_{z'} g_{\rho\gamma}(x',z') \Big\rangle \nn && - \Big\langle g_{\mu\nu}(x',z') \Big( \partial_{z} \mathcal{G}_{\mu\nu\rho\gamma}(x,z) - \frac{\partial }{\partial g^{\rho\gamma}(x,z)} \beta_{\mu\nu}[ g_{\mu\nu}(x,z);z] \Big)^{-1} \beta_{\rho\gamma}[ g_{\mu\nu}(x,z);z] \Big\rangle , \eqa
\end{small} 
which is analogous to the fluctuation-dissipation theorem away from equilibrium. 
Here, the Einstein tensor $G_{\mu\nu}(x,z) \sim \beta_{\mu\nu}[ g_{\mu\nu}(x,z);z]$ 
\cite{Nonperturbative_Wilson_RG, Einstein_Klein_Gordon_RG_Kim} 
would result in entropy production in the holographic RG flow.

All these thermodynamics perspectives motivate us to find an effective entropy functional in terms of the coupling field, expected to show its monotonicity during the RG evolution along the extradimension. 
%
%Firstly, we proposed an effective $\mathcal{C}_{eff}(z_{f})$ function Eq. (\ref{C_Function}), which corresponds to an onshell effective action of the bulk RG flow part. This first proposal is far from rigorous. However, this effective $\mathcal{C}_{eff}(z_{f})$ function appeared in the generalized fluctuation-dissiption theorem, Eq. (\ref{Fluctuation_Dissipation}) for the RG flow. More surprisingly, this $\mathcal{C}_{eff}(z_{f})$ function turns out to be identified with the environmental entropy, where the rate of this environmental entropy is nothing but the rate of heat dissipation in the medium, being responsible for an increase in entropy of the medium. 
%
Following Ref.\ \cite{Entropy_Production}, 
we discussed the entropy production during the RG flow. 
First, we introduced a probability distribution function for the coupling field, 
where the effective Fokker-Planck equation has been modified by the RG effective potential. 
Based on this probability distribution function, 
we proposed a microscopic definition for the entropy of the bulk system, 
and considered the RG evolution of the system entropy, 
resorting to the modified Fokker-Planck equation. 
This leads us to introduce the rate of heat dissipation in the medium, 
identified with the rate of environmental entropy. 
Combining these two contributions, 
we could find that the total entropy production rate 
is always positive after the ensemble averaging. 
This positive total entropy production rate confirms 
the monotonicity or irreversibility of the RG flow.

We would like to point out that our nonequilibrium thermodynamics perspectives 
for the monotonicity of 
the RG flow may have an interesting geometrical interpretation. 
It has been demonstrated that the holographic RG flow is given by the Ricci flow equation, 
where the extradimensional coordinate plays the role of time 
in the evolution of the geometry from UV to IR 
\cite{Holographic_Duality_V, Holographic_Duality_VI, Holographic_Duality_VII,Holographic_RG_Flow_Ricci_Flow_I, Holographic_RG_Flow_Ricci_Flow_II}. 
We recall that the general RG flow equation 
%\textcolor{red}{
can be made manifest
%}
%manifested 
in the level of an effective action 
with the introduction of an emergent extradimensional space, 
regarded to be emergent dual holography \cite{Nonperturbative_Wilson_RG_Disorder, Nonperturbative_Wilson_RG, Einstein_Klein_Gordon_RG_Kim, Einstein_Dirac_RG_Kim,RG_GR_Geometry_I_Kim, RG_GR_Geometry_II_Kim}. 
It has been also shown that the Ricci flow \cite{Ricci_Flow_0, Ricci_Flow_I, Ricci_Flow_II} 
is a gradient flow \cite{Ricci_Flow_III}, 
where the evolution of the induced metric in the ADM hypersurface is given by a gradient of a functional. 
Indeed, G. Perelman constructed the so-called ``entropy" functional 
and showed that the Ricci flow belongs to the gradient flow 
with positive definite metric, 
extremizing his entropy functional \cite{Ricci_Flow_III, Ricci_NLsM_Gradient_i, Ricci_NLsM_Gradient_ii}. 
He was able to show the monotonicity of the Ricci flow based on his entropy functional.
We speculate that our entropy functional constructed 
from the probability distribution function serves 
as a microscopic description
for the macroscopic thermodynamics entropy functional, 
analogous to Perelman's entropy functional 
\cite{Gibbs_Entropy_for_Perelman_I,Gibbs_Entropy_for_Perelman_II,Gibbs_Entropy_for_Perelman_III}. 
We hope to clarify this aspect in our future study.

Unfortunately, we could not reveal a clear connection from the monotonicity 
of the RG flow based on our holographic dual effective field theory 
to the holographic $c-$theorem of Refs.\ \cite{c_theorem_holography_i,c_theorem_holography_ii}. 
In the holographic $c-$theorem, the so-called holographic $c$-function has been constructed 
from geometry and shown to have monotonicity. 
It would be interesting to understand how these three frameworks, 
(1) our microscopic construction of the entropy functional, 
(2) the macroscopic description of the Perelman's entropy functional, 
and (3) the geometric construction of the holographic $c$-function, are related.

\begin{acknowledgments}
K.-S. Kim is supported by the Ministry of Education, Science, and Technology (NRF-2021R1A2C1006453 and NRF-2021R1A4A3029839) of the National Research Foundation of Korea (NRF) and by TJ Park Science Fellowship of the POSCO TJ Park Foundation.
S.R.~is supported by the National Science Foundation under 
Award No.\ DMR-2001181, and by a Simons Investigator Grant from
the Simons Foundation (Award No.~566116). This work is also supported by
the Gordon and Betty Moore Foundation through Grant
GBMF8685 toward the Princeton theory program. 
K.-S. Kim appreciates helpful discussions with A. Mitra, D. Mukherjee, M. Nishida, and Jae-Hyuk Oh.
\end{acknowledgments}

\appendix

\section{Derivation of the Fokker-Planck equation}

We recall the generating functional for the overdamped Langevin dynamics,
\begin{small}
\bqa && \mathcal{W} = \mathcal{N} \int_{x_{i}}^{x_{f}} D x(t) D p(t) D c(t) D \bar{c}(t) \exp\Big[- \int_{t_{i}}^{t_{f}} d t \Big\{ i p(t) \Big( \partial_{t} x(t) - \mu F[x(t)] \Big) + D p^{2}(t) + \bar{c}(t) \Big(\partial_{t} - \mu \partial_{x} F[x(t)]\Big) c(t) \Big\} \Big] . \nn \eqa
\end{small}
This generating functional indicates that the ``Hamiltonian" would be given by
%
%\bqa && [x(t),p(t)] = i . \eqa
%
\bqa && \mathcal{H} = - i p(t) \mu F[x(t)] + D p^{2}(t) \longrightarrow - \frac{\partial }{\partial x} \mu F(x) + D \frac{\partial^{2} }{\partial x^{2}} . \eqa
Hinted from
\bqa && \partial_{t} \Psi(x,t) = \Big( - \frac{\partial }{\partial x} \mu F(x) + D \frac{\partial^{2} }{\partial x^{2}} \Big) \Psi(x,t) , \eqa 
we obtain
\bqa && \partial_{t} p(x,t) = - \partial_{x} j(x,t) = - \partial_{x} [ (\mu F(x) - D \partial_{x}) p(x,t) ] , \eqa 
where $\Psi(x,t)$ is identified with $p(x,t)$.

%
%\subsection{The Wheeler–DeWitt equation approach}
%

Following this strategy, we derive the Fokker–Planck equation for the holographic dual effective field theory. The partition function is given by
\begin{small}
\bqa && \mathcal{Z}(z_{f}) = \int D \lambda(x,z) D \pi(x,z) D \bar{c}(x,z) D c(x,z) \exp\Big[ - N \int_{\Lambda_{uv}}^{z_{f}} d z \int d^{D} x \Big\{ i \pi(x,z) \Big( \partial_{z} \lambda(x,z) - \beta[\lambda(x,z);z] \Big) + \frac{\Gamma}{2} \pi^{2}(x,z) \nn && + \bar{c}(x,z) \Big(\partial_{z} - \frac{\partial \beta[\lambda(x,z);z]}{\partial \lambda(x,z)} \Big) c(x,z) + \mathcal{V}_{rg}[\lambda(x,z);z] \Big\} \Big] . \eqa  
\end{small}
This expression gives the ``Hamiltonian" as
%
%\bqa && [\lambda(x,z), \pi(x',z)] = i \delta^{(D)}(x-x') . \eqa
%
\bqa && \mathcal{H} = \mathcal{V}_{rg}[\lambda(x,z);z] - i \pi(x,z) \beta[\lambda(x,z);z] + \frac{\Gamma}{2} \pi^{2}(x,z) \longrightarrow \mathcal{V}_{rg}(\lambda,z) - \frac{\partial}{\partial \lambda} \beta(\lambda,z) + \frac{\Gamma}{2} \frac{\partial^{2}}{\partial \lambda^{2}} . \eqa
As a result, it is natural to propose the Fokker–Planck equation of the RG flow as
%
%\bqa && \partial_{z} \rho(\lambda,z) = \Big( \mathcal{V}_{rg}(\lambda,z) - \frac{\partial}{\partial \lambda} \beta(\lambda,z) + \frac{\Gamma}{2} \frac{\partial^{2}}{\partial \lambda^{2}} \Big) \Psi(\lambda,z) . \eqa
%
\bqa && \Big( \partial_{z} - \mathcal{V}_{rg}(\lambda,z) \Big) \rho(\lambda,z) = - \partial_{\lambda} \Big\{ \Big( \beta(\lambda,z) - \frac{\Gamma}{2} \partial_{\lambda} \Big) \rho(\lambda,z) \Big\} . \eqa
We emphasize that this Fokker-Planck equation of the RG flow is semiclassical, i.e., justified in the large $N$-limit.

\section{Superspace formulation}

In this appendix, we reformulate the holographic dual effective field theory in superspace, 
following Refs.\ \cite{MSR_Formulation_SUSY_i, MSR_Formulation_SUSY_ii, MSR_Formulation_SUSY_iii, MSR_Formulation_SUSY_iv, MSR_Formulation_SUSY_v, MSR_Formulation_SUSY_vi}. 
The superspace formulation shows transparently that both the boundary action and the RG-generated effective potential are two sources to break the previously introduced BRST symmetries, 
more precisely, $\mathcal{N} = 2$ supersymmetry. 
This point clarifies that the RG-generated effective potential is responsible for the entropy production during the RG flow.

First, we introduce a superfield,
\bqa && \Phi(x,z,\theta,\bar{\theta}) = \lambda(x,z) + \theta \bar{c}(x,z) + c(x,z) \bar{\theta} + \theta \bar{\theta} \pi(x,z) , \eqa
and its source field,
\bqa && J(x,z,\theta,\bar{\theta}) = T(x,z) + \theta \bar{G}(x,z) + G(x,z) \bar{\theta} + \theta \bar{\theta} \bar{T}(x,z) , \eqa 
respectively. 
Here, $\theta$ and $\bar{\theta}$ are Grassmann coordinates in superspace, 
where $\int_{\Lambda_{uv}}^{z_{f}} d z \int d^{D} x$ is replaced with $\int_{\Lambda_{uv}}^{z_{f}} d z \int d^{D} x d \bar{\theta} d \theta$.

%
%\subsection{BRST operators in the superspace formulation}
%

In this superspace formulation, the first two BRST charges $Q$ and $\bar{Q}$ are represented by
\bqa && Q \Phi(x,z,\theta,\bar{\theta}) = \Big(c(x,z) \frac{\delta }{\delta \lambda(x,z)} - \pi(x,z) \frac{\delta }{\delta \bar{c}(x,z)}\Big) \Big( \lambda(x,z) + \theta \bar{c}(x,z) + c(x,z) \bar{\theta} + \theta \bar{\theta} \pi(x,z) \Big) \nn && = c(x,z) + \theta \pi(x,z) = - \partial_{\bar{\theta}} \Big( \lambda(x,z) + \theta \bar{c}(x,z) + c(x,z) \bar{\theta} + \theta \bar{\theta} \pi(x,z) \Big) , \\
%
%. \eqa
%Accordingly, the second BRST charge $\bar{Q}$ is given by
%\bqa 
%
&& \bar{Q} \Phi(x,z,\theta,\bar{\theta}) = \Big\{\bar{c}(x,z) \frac{\delta }{\delta \lambda(x,z)} + \frac{2}{\Gamma} [\partial_{z} \bar{c}(x,z)] \frac{\delta }{\delta \pi(x,z)} + \Big( \pi(x,z) - \frac{2}{\Gamma} \partial_{z} \lambda(x,z) \Big) \frac{\delta }{\delta c(x,z)}\Big\} \Big( \lambda(x,z) + \theta \bar{c}(x,z) \nn && + c(x,z) \bar{\theta} + \theta \bar{\theta} \pi(x,z) \Big) = \bar{c}(x,z) + \bar{\theta} \Big( \pi(x,z) - \frac{2}{\Gamma} \partial_{z} \lambda(x,z) \Big) + \frac{2}{\Gamma} \theta \bar{\theta} \partial_{z} \bar{c}(x,z) \nn && = \Big(\partial_{\theta} - \frac{2}{\Gamma} \bar{\theta} \partial_{z} \Big) \Big( \lambda(x,z) + \theta \bar{c}(x,z) + c(x,z) \bar{\theta} + \theta \bar{\theta} \pi(x,z) \Big) , \eqa
respectively. The second two BRST charges $D$ and $\bar{D}$ are given by 
\bqa && D \Phi(x,z,\theta,\bar{\theta}) = \Big( \bar{c}(x,z) \frac{\delta }{\delta \lambda(x,z)} + \pi(x,z) \frac{\delta }{\delta c(x,z)} \Big) \Big( \lambda(x,z) + \theta \bar{c}(x,z) + c(x,z) \bar{\theta} + \theta \bar{\theta} \pi(x,z) \Big) \nn && = \bar{c}(x,z) + \bar{\theta} \pi(x,z) = \partial_{\theta} \Big( \lambda(x,z) + \theta \bar{c}(x,z) + c(x,z) \bar{\theta} + \theta \bar{\theta} \pi(x,z) \Big) , \\
%
%. \eqa
%The last BRST charge $\bar{D}$ is
%\bqa 
%
&& \bar{D} \Phi(x,z,\theta,\bar{\theta}) = \Big\{ c(x,z) \frac{\delta }{\delta \lambda(x,z)} + \frac{2}{\Gamma} [\partial_{z} c(x,z)] \frac{\delta }{\delta \pi(x,z)} - \Big( \pi(x,z) - \frac{2}{\Gamma} \partial_{z} \lambda(x,z) \Big) \frac{\delta }{\delta \bar{c}(x,z)} \Big\} \Big( \lambda(x,z) + \theta \bar{c}(x,z) \nn && + c(x,z) \bar{\theta} + \theta \bar{\theta} \pi(x,z) \Big) = c(x,z) + \theta \Big( \pi(x,z) - \frac{2}{\Gamma} \partial_{z} \lambda(x,z) \Big) + \theta \bar{\theta} \frac{2}{\Gamma} [\partial_{z} c(x,z)] \nn && = - \Big( \partial_{\bar{\theta}} + \frac{2}{\Gamma} \theta \partial_{z} \Big) \Big( \lambda(x,z) + \theta \bar{c}(x,z) + c(x,z) \bar{\theta} + \theta \bar{\theta} \pi(x,z) \Big) , \eqa
respectively.

Anti-commutation relations of these BRST charges are given by
\bqa && [Q, \bar{Q}]_{+} = Q \bar{Q} + \bar{Q} Q = - \partial_{\bar{\theta}_{a}} \Big(\partial_{\theta_{a}} - \frac{2}{\Gamma_{a}} \bar{\theta}_{a} \partial_{z} \Big) - \Big( \partial_{\theta_{a}} - \frac{2}{\Gamma_{a}} \bar{\theta}_{a} \partial_{z} \Big) \partial_{\bar{\theta}_{a}} = \frac{2}{\Gamma_{a}} \partial_{z} , \\ && [D, \bar{D}]_{+} = D \bar{D} + \bar{D} D = - \partial_{\theta_{a}} \Big(\partial_{\bar{\theta}_{a}} + \frac{2}{\Gamma_{a}} \theta_{a} \partial_{z} \Big) - \Big(\partial_{\bar{\theta}_{a}} + \frac{2}{\Gamma_{a}} \theta_{a} \partial_{z} \Big) \partial_{\theta_{a}} = - \frac{2}{\Gamma_{a}} \partial_{z} , \eqa
both of which lead to the translation operator along the extra dimension.

%
%\subsection{Effective partition function in the superspace formulation}
%

Based on these constructions, we rewrite the partition function as follows
\bqa && Z(z_{f}) = \int D \psi_{\sigma}(x,z_{f}) D \Phi(x,z,\theta,\bar{\theta}) \exp\Big[ - \int d^{D} x \Big\{ \mathcal{L}[\psi_{\sigma}(x,z_{f});\lambda(x,z_{f});z_{f}] + \frac{N}{2 \Gamma}[\lambda(x,\Lambda_{uv}) - \bar{\lambda}(\Lambda_{uv})]^{2} \Big\} \nn && - N \int_{\Lambda_{uv}}^{z_{f}} d z \int d^{D} x d \bar{\theta} d \theta \Big\{ \frac{\Gamma}{2} D \Phi(x,z,\theta,\bar{\theta}) \bar{D} \Phi(x,z,\theta,\bar{\theta}) + \mathcal{V}_{rg}[\Phi(x,z,\theta,\bar{\theta});z] + J(x,z,\theta,\bar{\theta}) \Phi(x,z,\theta,\bar{\theta}) \Big\} \nn && - N \int_{\Lambda_{uv}}^{z_{f}} d z \int d^{D} x \mathcal{V}_{rg}[\lambda(x,z);z] \Big] . \eqa  
The kinetic energy is checked out as 
\bqa && \int d^{D} x d \bar{\theta} d \theta \frac{\Gamma}{2} D \Phi(x,z,\theta,\bar{\theta}) \bar{D} \Phi(x,z,\theta,\bar{\theta}) \nn && =  \int d^{D} x d \bar{\theta} d \theta \frac{\Gamma}{2} \Big\{ \bar{c}(x,z) + \bar{\theta} \pi(x,z) \Big\} \Big\{ c(x,z) + \theta \Big( \pi(x,z) - \frac{2}{\Gamma} \partial_{z} \lambda(x,z) \Big) + \theta \bar{\theta} \frac{2}{\Gamma} [\partial_{z} c(x,z)] \Big\} \nn && = \int d^{D} x \Big\{ \pi(x,z) \partial_{z} \lambda(x,z) - \frac{\Gamma}{2} \pi^{2}(x,z) + \bar{c}(x,z) \partial_{z} c(x,z) \Big\} . \eqa
The RG $\beta$ function is generated by the effective potential
\bqa && \mathcal{V}_{rg}[\Phi(x,z,\theta,\bar{\theta});z] = - \frac{1}{N} \ln \int_{\Lambda(z)} D \psi_{\sigma}(x,z) \exp\Big\{ - \int d^{D} x d \bar{\theta} d \theta \mathcal{L}[\psi_{\sigma}(x,z);\Phi(x,z,\theta,\bar{\theta});z]\Big\} \eqa
in the superspce. The source-field coupling action is 
\bqa && \int d^{D} x d \bar{\theta} d \theta J(x,z,\theta,\bar{\theta}) \Phi(x,z,\theta,\bar{\theta}) \nn && = \int d^{D} x d \bar{\theta} d \theta \Big(T(x,z) + \theta \bar{G}(x,z) + G(x,z) \bar{\theta} + \theta \bar{\theta} \bar{T}(x,z)\Big) \Big( \lambda(x,z) + \theta \bar{c}(x,z) + c(x,z) \bar{\theta} + \theta \bar{\theta} \pi(x,z)\Big) \nn && = \int d^{D} x \Big(\bar{T}(x,z) \lambda(x,z) + T(x,z) \pi(x,z) + \bar{G}(x,z) c(x,z) + \bar{c}(x,z) G(x,z) \Big) . \eqa


\begin{thebibliography}{9}
\bibitem{c_theorem} A. B. Zamolodchikov, \textit{Irreversibility of the Flux of the Renormalization Group in a 2D Field Theory}, JETP Lett. \textbf{43}, 730 (1986) [Pisma Zh. Eksp. Teor. Fiz.43,565(1986)].
\bibitem{a_theorem} J. L. Cardy, \textit{Is There a c Theorem in Four-Dimensions?}, Phys. Lett. B \textbf{215}, 749 (1988).
\bibitem{a_f_theorem_i} Z. Komargodski and A. Schwimmer, \textit{On Renormalization Group Flows in Four Dimensions}, JHEP \textbf{12}, 099 (2011); [arXiv:1107.3987].
\bibitem{f_theorem_SUSY} D. L. Jafferis, I. R. Klebanov, S. S. Pufu, and B. R. Safdi, \textit{Towards the F-Theorem: N = 2 Field Theories on the Three-Sphere}, JHEP \textbf{06}, 102 (2011); [arXiv:1103.1181].
\bibitem{f_theorem_noSUSY} I. R. Klebanov, S. S. Pufu, and B. R. Safdi, \textit{F-Theorem without Supersymmetry}, JHEP \textbf{10}, 038 (2011); [arXiv:1105.4598].
\bibitem{a_f_theorem_ii} S. Giombi and I. R. Klebanov, \textit{Interpolating between a and F}, JHEP \textbf{03}, 117 (2015); [arXiv:1409.1937].
\bibitem{Entanglement_Entropy_Review_RMP} T. Nishioka, \textit{Entanglement entropy: Holography and renormalization group}, Rev. Mod. Phys. \textbf{90}, 035007 (2018).
%
%[12] H. Liu and M. Mezei, A Refinement of entanglement entropy and the number of degrees of freedom, JHEP 04 (2013) 162, [arXiv:1202.2070].
%
\bibitem{a_f_theorem_EE_i} H. Casini and M. Huerta, \textit{On the RG running of the entanglement entropy of a circle}, Phys. Rev. D \textbf{85}, 125016 (2012); [arXiv:1202.5650].
\bibitem{a_f_theorem_EE_ii} H. Casini and M. Huerta, \textit{A Finite entanglement entropy and the c-theorem}, Phys. Lett. B \textbf{600}, 142 (2004); [hep-th/0405111].
\bibitem{a_f_theorem_EE_iii} H. Casini, M. Huerta, and R. C. Myers, \textit{Towards a derivation of holographic entanglement entropy}, JHEP \textbf{05}, 036 (2011); [arXiv:1102.0440].
\bibitem{a_f_theorem_EE_iv} H. Casini, E. Test´e, and G. Torroba, \textit{Markov Property of the Conformal Field Theory Vacuum and the a Theorem}, Phys. Rev. Lett. \textbf{118}, 261602 (2017); [arXiv:1704.01870].
\bibitem{a_f_theorem_EE_v} H. Casini, I. S. Landea, and G. Torroba, \textit{The g-theorem and quantum information theory}, JHEP \textbf{10}, 140 (2016); [arXiv:1607.00390].
\bibitem{c_theorem_holography_i} R. C. Myers and A. Sinha, \textit{Seeing a c-theorem with holography}, Phys. Rev. D \textbf{82}, 046006 (2010); [arXiv:1006.1263].
\bibitem{c_theorem_holography_ii} R. C. Myers and A. Sinha, \textit{Holographic c-theorems in arbitrary dimensions}, JHEP \textbf{01}, 125 (2011); [arXiv:1011.5819]. 
\bibitem{MSR_Formulation_SUSY_i} G. Parisi and N. Sourlas, \textit{Random Magnetic Fields, Supersymmetry, and Negative Dimensions}, Phys. Rev. Lett. \textbf{43}, 744 (1979).
\bibitem{MSR_Formulation_SUSY_ii} S. Chaturvedi, A. K. Kapoor, and V. Srinivasan, \textit{Ward Takahashi identities and fluctuation-dissipation theorem in a superspace formulation of the Langevin equation}, Z. Phys. B \textbf{57}, 249 (1984).
\bibitem{MSR_Formulation_SUSY_iii} K. Mallick, M. Moshe, and H. Orland, \textit{A field-theoretic approach to non-equilibrium work identities}, J. Phys. A: Math. Gen. \textbf{44}, 095002 (2011).
\bibitem{MSR_Formulation_SUSY_iv} J. Zinn-Justin, \textit{Quantum Field Theory and Critical Phenomena}, International Series of Monographs on Physics (Clarendon Press, London, 2002).
\bibitem{MSR_Formulation_SUSY_v} I. Ovchinnikov, \textit{Introduction to Supersymmetric Theory of Stochastics}, Entropy \textbf{18}, 108 (2016).
\bibitem{MSR_Formulation_SUSY_vi} Piotr Surowka and Piotr Witkowski, \textit{Symmetries in the path integral formulation of the Langevin dynamics}, Phys. Rev. E \textbf{98}, 042140 (2018).
\bibitem{MERA} G. Vidal, \textit{Class of Quantum Many-Body States That Can Be Efficiently Simulated}, Phys.\ Rev.\ Lett.\ \textbf{101}, 110501 (2008). 
\bibitem{Jarzynski_i} C. Jarzynski, \textit{Nonequilibrium Equality for Free Energy Differences}, Phys. Rev. Lett. \textbf{78}, 2690 (1997).
\bibitem{Jarzynski_ii} C. Jarzynski, \textit{Equilibrium free-energy differences from nonequilibrium measurements: A master-equation approach}, Phys. Rev. E \textbf{56}, 5018 (1997).
\bibitem{Crooks_i} G. E. Crooks, \textit{Nonequilibrium Measurements of Free Energy Differences for Microscopically Reversible Markovian Systems}, J. Stat. Phys. \textbf{90}, 1481 (1998).
\bibitem{Crooks_ii} G. E. Crooks, \textit{Entropy production fluctuation theorem and the nonequilibrium work relation for free energy differences}, Phys. Rev. E \textbf{60}, 2721 (1999).
\bibitem{Crooks_iii} G. E. Crooks, \textit{Path-ensemble averages in systems driven far from equilibrium}, Phys. Rev. E \textbf{61}, 2361 (2000).
\bibitem{Entropy_Production} Udo Seifert, \textit{Entropy Production along a Stochastic Trajectory and an Integral Fluctuation Theorem}, Phys. Rev. Lett. \textbf{95}, 040602 (2005).
\bibitem{Schwinger_Keldysh_Symmetries_i} F. M. Haehl, R. Loganayagam, and M. Rangamani, \textit{The fluid manifesto: emergent symmetries, hydrodynamics, and black holes}, JHEP \textbf{01}, 184 (2016).
\bibitem{Schwinger_Keldysh_Symmetries_ii} F. M. Haehl, R. Loganayagam, and M. Rangamani, \textit{Schwinger-Keldysh formalism. Part I: BRST symmetries and superspace}, JHEP \textbf{06}, 069 (2017).
\bibitem{Schwinger_Keldysh_Symmetries_iii} F. M. Haehl, R. Loganayagam, and M. Rangamani, \textit{Schwinger-Keldysh formalism. Part II: thermal equivariant cohomology}, JHEP \textbf{06}, 070 (2017).
\bibitem{Schwinger_Keldysh_Symmetries_iv} P. Glorioso and H. Liu, \textit{The second law of thermodynamics from symmetry and unitarity}, arXiv:1612.07705 [hep-th].
\bibitem{Schwinger_Keldysh_Symmetries_v} P. Glorioso, M. Crossley, and H. Liu, \textit{Effective field theory of dissipative fluids (II): classical limit, dynamical KMS symmetry and entropy current}, JHEP \textbf{09}, 096 (2017).
\bibitem{Schwinger_Keldysh_Symmetries_vi} K. Jensen, N. Pinzani-Fokeeva, and A. Yarom, \textit{Dissipative hydrodynamics in superspace}, JHEP \textbf{09}, 127 (2018).
\bibitem{TQFT_Witten_Type} D. Birmingham, M. Blau, M. Rakowski, and G. Thompson, \textit{Topological field theory}, Phys. Rep. \textbf{209}, 129 (1991).
\bibitem{MSR_Formulation_i} P. C. Martin, E. D. Siggia, and H. A. Rose, \textit{Statistical Dynamics of Classical Systems}, Phys. Rev. A \textbf{8}, 423 (1973).
\bibitem{MSR_Formulation_ii} H. K. Janssen, \textit{On a Lagrangean for Classical Field Dynamics and Renormalization Group Calculations of Dynamical Critical Properties}, Z. Phys. B \textbf{23} 377 (1976).
\bibitem{MSR_Formulation_iii} C. de Dominicis and L. Peliti, \textit{Field-theory renormalization and critical dynamics above $T_{c}$: Helium, antiferromagnets, and liquid-gas systems}, Phys. Rev. B \textbf{18}, 353 (1978).
\bibitem{NEQ_textbook} A. Kamenev, \textit{Field Theory of Non-Equilibrium Systems}, (Cambridge University Press, Cambridge, 2011).
\bibitem{Nonperturbative_Wilson_RG_Disorder} Ki-Seok Kim, \textit{Beyond quantum chaos in emergent dual holography}, Phys. Rev. D \textbf{106}, 126014 (2022).
\bibitem{Nonperturbative_Wilson_RG} Ki-Seok Kim, Shinsei Ryu, and Kanghoon Lee, \textit{Emergent dual holographic description as a nonperturbative generalization of the Wilsonian renormalization group}, Phys. Rev. D \textbf{105}, 086019 (2022).
\bibitem{Einstein_Klein_Gordon_RG_Kim} Ki-Seok Kim and Shinsei Ryu, \textit{Entanglement transfer from quantum matter to classical geometry in an emergent holographic dual description of a scalar field theory}, JHEP05(2021)260, https://doi.org/10.1007/JHEP05(2021)260.
\bibitem{Einstein_Dirac_RG_Kim} Ki-Seok Kim, \textit{Emergent dual holographic description for interacting Dirac fermions in the large $N$ limit}, Phys. Rev. D \textbf{102}, 086014 (2020).
\bibitem{RG_GR_Geometry_I_Kim} Ki-Seok Kim, \textit{Geometric encoding of renormalization group $\beta$-functions in an emergent holographic dual description}, Phys. Rev. D \textbf{102}, 026022 (2020).
\bibitem{RG_GR_Geometry_II_Kim} Ki-Seok Kim, \textit{Emergent geometry in recursive renormalization group transformations}, Nucl. Phys. B \textbf{959}, 115144 (2020).
\bibitem{Kondo_Holography_Kim} Ki-Seok Kim, Suk Bum Chung, Chanyong Park, and Jae-Ho Han, \textit{A non-perturbative field theory approach for the Kondo effect: Emergence of an extra dimension and its implication for the holographic duality conjecture}, Phys. Rev. D \textbf{99}, 105012 (2019).
\bibitem{Kitaev_Entanglement_Entropy_Kim} K.-S. Kim, M. Park, J. Cho, and C. Park, \textit{An emergent geometric description for a topological phase transition in the Kitaev superconductor model}, Phys. Rev. D \textbf{96}, 086015 (2017).
\bibitem{RG_Holography_First_Kim} K.-S. Kim and C. Park, \textit{Emergent geometry from field theory: Wilson's renormalization group revisited}, Phys. Rev. D \textbf{93}, 121702 (2016).
\bibitem{Emergent_AdS2_BH_RG} Ki-Seok Kim, Mitsuhiro Nishida, and Yoonseok Choun, \textit{Renormalization group flow to effective quantum mechanics at IR in an emergent dual holographic description for spontaneous chiral symmetry breaking}, Phys. Rev. D \textbf{107}, 066004 (2023).
\bibitem{QFT_textbook} Michael E. Peskin and Daniel V. Schroeder, \textit{An Introduction To Quantum Field Theory}, (CRC Press. Taylor and Francis Group, New York, 1995).
\bibitem{Holographic_Duality_I} J. M. Maldacena, \textit{The Large $N$ Limit of Superconformal Field Theories and Supergravity}, Int. J. Theor. Phys. \textbf{38}, 1113 (1999).
\bibitem{Holographic_Duality_II} S. S. Gubser, I. R. Klebanov, and A. M. Polyakov, textit{Gauge Theory Correlators from Non-Critical String Theory}, Phys. Lett. B \textbf{428}, 105 (1998).
\bibitem{Holographic_Duality_III} E. Witten, \textit{Anti De Sitter Space And Holography}, Adv. Theor. Math. Phys. \textbf{2}, 253 (1998).
\bibitem{Holographic_Duality_IV} O. Aharony, S. S. Gubser, J. Maldacena, H. Ooguri, and Y. Oz, \textit{Large $N$ Field Theories, String Theory and Gravity}, Phys. Rep. \textbf{323}, 183 (2000).
\bibitem{Holographic_Duality_V} M. Bianchi, D. Z. Freedman, and K. Skenderis, \textit{Holographic renormalization}, Nucl. Phys. B \textbf{631}, 159 (2002).
\bibitem{Holographic_Duality_VI} J. de Boer, E. P. Verlinde, and H. L. Verlinde, \textit{On the Holographic Renormalization Group}, JHEP \textbf{08}, 003 (2000).
\bibitem{Holographic_Duality_VII} E. P. Verlinde and H. L. Verlinde, \textit{RG flow, gravity and the cosmological constant}, JHEP \textbf{05}, 034 (2000).
\bibitem{Higher_Spin_Gauge_Theory_I} C. Fronsdal, \textit{Massless fields with integer spin}, Phys. Rev. D \textbf{18}, 3624 (1978).
\bibitem{Higher_Spin_Gauge_Theory_II} E. S. Fradkin and M. A. Vasiliev, \textit{On the gravitational interaction of massless higher-spin fields}, Phys. Lett. B \textbf{189}, 89 (1987); E. S. Fradkin and M. A. Vasiliev, \textit{Cubic interaction in extended theories of massless higher-spin fields}, Nucl. Phys. B \textbf{291}, 141 (1987).
\bibitem{Higher_Spin_Gauge_Theory_III} I. R. Klebanov and A. M. Polyakov, \textit{AdS dual of the critical O(N) vector model}, Phys. Lett. B \textbf{550}, 213 (2002).
\bibitem{Higher_Spin_Gauge_Theory_IV} X. Bekaert, N. Boulanger, and P. A. Sundell, \textit{How higher-spin gravity surpasses the spin-two barrier}, Rev. Mod. Phys. \textbf{84}, 987 (2012).
\bibitem{Holography_Higher_Spin_RG_I} E. Mintun and J. Polchinski, \textit{Higher spin holography, RG, and the light cone}, arXiv:1411.3151.
\bibitem{Holography_Higher_Spin_RG_II} R. G. Leigh, O. Parrikar, and A. B. Weiss, \textit{Holographic geometry of the renormalization group and higher spin symmetries}, Phys. Rev. D \textbf{89}, 106012  (2014).
\bibitem{Holography_Higher_Spin_RG_III} R. G. Leigh, O. Parrikar, and A. B. Weiss, \textit{Exact renormalization group and higher-spin holography}, Phys. Rev. D \textbf{91}, 026002 (2015).
\bibitem{Higher_Spin_Gauge_Theory_V} Ofer Aharony, Shai M. Chester, and Erez Y. Urbach, \textit{A derivation of AdS/CFT for vector models}, JHEP \textbf{03} 208 (2021).
\bibitem{TTbar_Deformation} J. Cardy, \textit{The $T\bar{T}$ deformation of quantum field theory as random geometry}, JHEP \textbf{10} (2018) 186 [arXiv:1801.06895].
\bibitem{ADM_Hamiltonian_Formulation} R. Arnowitt, S. Deser, and C. W. Misner, \textit{Dynamical Structure and Definition of Energy in General Relativity}, Phys. Rev. \textbf{116}, 1322 (1959).
\bibitem{DeWitt_Metric} Bryce S. DeWitt, \textit{Quantum Theory of Gravity. I. The Canonical Theory}, Phys. Rev. \textbf{160}, 1113 (1967).
\bibitem{Holographic_RG_Flow_Ricci_Flow_I} E. Kiritsis, W. Li, and F. Nitti, \textit{Holographic RG flow and the Quantum Effective Action}, Fortsch. Phys. \textbf{62}, 389-454 (2014) doi:10.1002/prop.201400007 [arXiv:1401.0888].
\bibitem{Holographic_RG_Flow_Ricci_Flow_II} S. Jackson, R. Pourhasan, and H. Verlinde, \textit{Geometric RG Flow}, arXiv:1312.6914.
\bibitem{Ricci_Flow_0} R. S. Hamilton, \textit{Hamilton : Three-manifolds with positive Ricci curvature}, J. Diff. Geom. \textbf{17}, 255 (1982).
\bibitem{Ricci_Flow_I} D. Friedan, \textit{Nonlinear Models in $2 + \varepsilon$ Dimensions}, Phys. Rev. Lett. \textbf{45}, 1057 (1980); D. Friedan, \textit{Nonlinear models in $2 + \varepsilon$ dimensions}, Ann. Phys. \textbf{163}, 318 (1985).
\bibitem{Ricci_Flow_II} B. Chow and D. Knopf, \textit{The Ricci Flow: An Introduction}, Math. Surveys and Monographs \textbf{110}, Am. Math. Soc. (2004)
\bibitem{Ricci_Flow_III} G. Perelman, \textit{The entropy formula for the Ricci flow and its geometric applications}, arXiv:math/0211159v1; G. Perelman, \textit{Ricci flow with surgery on three-manifolds}, arXiv:math/0303109v1; G. Perelman, \textit{Finite extinction time for the solutions to the Ricci flow on certain three-manifolds}, arXiv:math/0307245v1.
\bibitem{Ricci_NLsM_Gradient_i} T. Oliynyk, V. Suneeta, and E. Woolgar, \textit{A gradient flow for worldsheet nonlinear sigma models}, Nucl. Phys. B \textbf{739}, 441 (2006).
\bibitem{Ricci_NLsM_Gradient_ii} A. A. Tseytlin, \textit{Sigma model renormalization group flow, ``central charge" action, and Perelman’s entropy}, Phys. Rev. D \textbf{75}, 064024 (2007).
\bibitem{Gibbs_Entropy_for_Perelman_I} Xiang-Dong Li, \textit{Perelman's entropy formula for the Witten Laplacian on Riemannian manifolds via Bakry-Emery Ricci curvature}, Math. Ann. \textbf{353} (2012), 403.
\bibitem{Gibbs_Entropy_for_Perelman_II} Xiang-Dong Li, \textit{From the Boltzmann $H$-theorem to Perelman's $W$-entropy formula for the Ricci flow}, arXiv:1303.5193 [math.DG], https://doi.org/10.48550/arXiv.1303.5193.
\bibitem{Gibbs_Entropy_for_Perelman_III} Xiang-Dong Li, \textit{Perelman's $W$-entropy for the Fokker–Planck equation over complete Riemannian manifolds}, Bull. Sci. math. \textbf{135} (2011) 871.
\end{thebibliography}
\end{document}